\documentclass[aps,onecolumn,11pt,floatfix,altaffilletter,tightenlines,
	showpacs,showkeys,notitlepage,nofootinbib]{revtex4-1}

\usepackage[colorlinks=true,citecolor=blue,linkcolor=blue]{hyperref}
\usepackage{amsmath,amssymb}
\usepackage{dsfont}
\usepackage{relsize}
\usepackage{slashed}
\usepackage{mathtools}
\usepackage{graphicx}             
\usepackage{url}
\usepackage{multirow}
\usepackage[dvipsnames]{xcolor}
\usepackage{csquotes}
\usepackage{xspace}
\usepackage{nicefrac}
\usepackage{titletoc}
\usepackage{siunitx}
\usepackage[capitalize]{cleveref}
\usepackage[caption=false]{subfig}
\usepackage{ifthen}

\contentsmargin{1.8em}
\titlecontents{section}
	[2.3em]
	{\addvspace{0.8em}\large}
	{\contentslabel[\thecontentslabel\ifthenelse{\equal{\thecontentslabel}{}}{}{.}]{2.3em}}
	{}
	{\titlerule*[0.7pc]{.}\contentspage}

\makeatletter
\def\l@subsection#1#2{}
\makeatother

\allowdisplaybreaks

\bibpunct{[}{]}{,}{n}{}{,}

\newcommand{\del}{\partial}
\newcommand{\dd}{\mbox{d}}

\DeclareMathOperator{\tr}{Tr}

\DeclarePairedDelimiter\chevron{\langle}{\rangle}
\DeclarePairedDelimiter\absVal{\vert}{\vert}

\newcommand{\I}{\mathrm{i}}
\newcommand{\E}{\mathrm{e}}
\newcommand{\ie}{i.e.~}
\newcommand{\eg}{e.g.~}
\newcommand{\cf}{cf.~}
\newcommand{\hc}{\text{h.c.\xspace}}
\newcommand{\tinytext}[1]{\text{\tiny{#1}}}
\newcommand{\fineq}[1]{\;{#1}}

\newcommand{\symhspace}[2]{\hspace{#1}#2\hspace{#1}}

\newcommand{\Lagn}{\mathcal{L}}
\newcommand{\unitOp}[1][]{\mathds{1}_{#1}}
\newcommand{\suN}[1]{\text{SU(}#1\text{)}}
\newcommand{\uN}[1]{\text{U(}#1\text{)}}
\newcommand{\oN}[1]{\text{O(}#1\text{)}}
\newcommand{\Nc}{n_c}
\newcommand{\Nf}{n_f}
\newcommand{\MPl}{M_\tinytext{Pl}}
\newcommand{\geff}{g_\star}
\newcommand{\heff}{h_\star}
\newcommand{\JB}{J_\tinytext{B}}
\newcommand{\JBF}{J_\tinytext{B,F}}
\newcommand{\JF}{J_\tinytext{F}}
\newcommand{\IB}{I_\tinytext{B}}

\newcommand{\MF}{\Psi}
\newcommand{\etap}{{\eta^\prime}}
\newcommand{\phiCl}{\bar{\phi}}
\newcommand{\PhiCl}{\bar{\Phi}}
\newcommand{\propTree}{\Delta}
\newcommand{\propCl}{\bar{G}}
\newcommand{\propPhys}{\hat{G}}
\newcommand{\sigCl}{\bar{\sigma}}
\newcommand{\sigMinT}{\bar{\sigma}_\text{min}}
\newcommand{\sigVEV}{v_\sigma}
\newcommand{\Tglue}{T_\text{glue}}

\newcommand{\Vgen}[2]{V_{#1}^\tinytext{#2}}
\newcommand{\Veff}[1]{\Vgen{\text{eff}}{#1}}
\newcommand{\Vzero}[1]{\Vgen{0}{#1}}
\newcommand{\VCW}[1]{\Vgen{\tinytext{CW}}{#1}}
\newcommand{\VFT}[1]{\Vgen{\tinytext{FT}}{#1}}
\newcommand{\Vtree}[1]{\Vgen{\text{tree}}{#1}}
\newcommand{\VCJT}[1]{\Vgen{\tinytext{CJT}}{#1}}


\begin{document}

\title{Observational prospects for gravitational waves\\from hidden or dark chiral phase transitions}

\author{Alexander J. Helmboldt$^{1}$} \email{alexander.helmboldt@mpi-hd.mpg.de}
\author{Jisuke Kubo$^{1,2\,}$} \email{jisuke.kubo@mpi-hd.mpg.de}
\author{Susan van der Woude$^{1}$}   \email{susan@mpi-hd.mpg.de}
\affiliation{%
	$^1$Max-Planck-Institut f\"ur Kernphysik, 69117~Heidelberg, Germany\\
	$^2$Department of Physics, University of Toyama, 3190 Gofuku, Toyama 930-8555, Japan%
}

\begin{abstract}

\noindent
We study the gravitational wave (GW) signature of first-order chiral phase transitions ($\chi$PTs) in strongly interacting hidden or dark sectors.
We do so using several effective models  in order to reliably capture the relevant non-perturbative dynamics.
This approach allows us to explicitly calculate key quantities characterizing the $\chi$PT without having to resort to rough estimates.
Most importantly, we find that the transition's inverse duration $\beta$ normalized to the Hubble parameter $H$ is at least two orders of magnitude larger than typically assumed in comparable scenarios, namely, $\beta/H\gtrsim\mathcal{O}(\num{e4})$.
The obtained GW spectra then suggest that signals from hidden $\chi$PTs occurring at around \SI{100}{MeV} might be in reach of LISA, while DECIGO and BBO may detect a stochastic GW background associated with transitions between roughly \SI{1}{GeV} and \SI{10}{TeV}. 
Signatures of transitions at higher temperatures are found to be outside the range of any currently proposed experiment. 
Even though predictions from different effective models are qualitatively similar, we find that they may vary considerably from a quantitative point of view, which highlights the need for true first-principle calculations such as lattice simulations.
\end{abstract}

\maketitle
{
	\hypersetup{linkcolor=black}
	\tableofcontents
	\makeatletter
	\let\toc@pre\relax
	\let\toc@post\relax
	\makeatother
}

\clearpage

\section{Introduction}
\label{sec:intro}
\noindent
Since their first direct detection on Earth \cite{Abbott2016,TheLIGOScientific2017,GBM2017}, gravitational waves (GWs) have provided a new and complementary approach towards observational astrophysics.
Not only do they allow to study violent transient phenomena like black hole or neutron star mergers, but they also offer a unique way to probe the physics of the early Universe.
In particular, it is well known that first-order cosmic phase transitions (PTs) can create a stochastic background of gravitational waves \cite{Witten1984,Hogan1984,Hogan1986b,Turner1990a,Kamionkowski1993,Grojean2007}.
Even though the Standard Model (SM) of particle physics does not feature such a PT \cite{Kajantie1996a,Kajantie1996b,Aoki2006a,Aoki2006b,Bhattacharya2014}, they \textit{are} predicted by a plethora of beyond-the-SM (BSM) theories.
For instance, successful electroweak baryogenesis typically requires a strong first-order \textit{electroweak} phase transition (EWPT), which can \eg be obtained in models with an augmented scalar particle content \cite{Morrissey2012a}.
Of course, gravitational waves may also be produced during phase transitions in dark or \textendash\ more generally \textendash\ hidden sectors.
Correspondingly, studies investigating gravitational wave signatures of a wide range of new physics scenarios have received a lot of attention in the recent years, see \eg \cite{Schwaller2015,Jaeckel2016a,Dev2016,Hashino2017,Baldes2017,Tsumura2017,Aoki2017,%
Croon2018,Okada2018a,Baldes2018,Ellis:2018mja,Prokopec2019,Bai2019,Brdar2019,%
Mazumdar2018,Miura2018,Addazi2018a,Marzo2018,Breitbach2018,Fairbairn2019a,Ellis:2019oqb}.

In the present article, we focus on BSM theories with a new confining gauge force $\suN{\Nc}$ acting on $\Nf$ flavors of hidden fermions in the fundamental representation.
Similar to the situation in quantum chromodynamics (QCD), strong (\ie non-perturbative) dynamics are anticipated to spontaneously break the hidden counterpart of chiral symmetry in such models.
Unlike in real-world QCD,%
\footnote{However, the crossover transition predicted for real-world QCD can influence the spectrum of inflationary gravitational waves \cite{Seto2003,Kuroyanagi2008,Schettler2010a,Schettler2010b,Saikawa2018}.}
the corresponding chiral phase transition can, however, be strong and of first order.
It is the gravitational wave background associated with such a transition that we will be interested in, \cf \cite{Schwaller2015}.

Hidden sectors of the above described type are particularly attractive for several reasons.
They are, for instance, featured in various models of dark matter (DM) like strongly interacting massive particle (SIMP) DM \cite{Hochberg2014,Hochberg2015}, or composite DM as discussed in \eg Refs.~\cite{Davoudiasl2012,Petraki2013,Zurek2014,Bai2014,%
Appelquist2013,Appelquist2014,Appelquist2015}.
Similarly, they are introduced in Twin Higgs models \cite{Chacko2006,Chacko2006a}, Hidden Valley theories \cite{Strassler2007} or vector-like confinement models \cite{Kilic2010}.
Lastly, strongly coupled hidden sectors commonly occur in theories based on classical scale invariance, since they offer one of the two%
\footnote{The other one being the (perturbative) Coleman-Weinberg mechanism \cite{Coleman1973}.}
known possibilities to dynamically generate a scale in a previously scaleless setup.
Several realistic classically conformal SM extensions along those lines were suggested \eg in Refs.~\cite{Hur2011,Heikinheimo2013,Holthausen2013,Kubo2014,Kubo2015b,Hatanaka2016}.
Reflecting this rich variety of applications, well-motivated scenarios exist at vastly different scales ranging from \SI{100}{MeV} to above \SI{10}{TeV}, see Ref.~\cite{Schwaller2015} for an overview.

Given the presented appealing physics case for BSM theories with new confining gauge forces, it is desirable to quantify gravitational wave signatures from a potential hidden chiral phase transition, in order to assess the prospects for their observation.
Due to the problem's non-perturbative nature, the corresponding analyses generally prove difficult.
Accordingly, no true first-principle calculations exist yet for the setup under consideration.%
\footnote{The bubble nucleation rate in a first-order EWPT was computed using first-principle lattice methods in Ref.~\cite{Moore2001a}. Similar calculations were performed in purely scalar theories \cite{Moore2001}. Based on some of the aforementioned lattice results and employing dimensionally reduced effective field theory, the authors of Ref.~\cite{Gould:2019qek} recently presented a non-perturbative analysis of the GW spectrum associated with a first-order EWPT.}
There are, however, various ways to proceed.
For a start, one may simply estimate the relevant quantities based on certain assumptions and/or general arguments.
For instance, the inverse duration of the phase transition $\beta$, which is a key parameter in understanding the associated GW spectrum, is commonly estimated by \cite{Hogan1984,Hogan1986b}
\begin{align}
	\frac{\beta}{H} \approx 4\log\frac{\MPl}{T_n} \approx \mathcal{O}(100) \fineq{,}
	\label{eq:intro:beta}
\end{align}
where $H$ is the Hubble parameter at the time of the transition, $\MPl$ is the (reduced) Planck mass and the second relation holds for a wide range of transition temperatures $T_n$.
Analyses of GW signals based on the estimate of \cref{eq:intro:beta} can \eg be found in Refs.~\cite{Caprini2010,Binetruy2012,Schwaller2015}.

Here, we will follow a different approach.
Namely, we will model the dynamics in the strongly coupled hidden sector by employing various effective theories, which are known to work well in capturing non-perturbative aspects of the low-energy regime of real-world QCD.
Thus we will be able to \textit{explicitly} calculate relevant quantities characterizing the hidden chiral phase transition such as the bubble nucleation rate, the transition temperature or $\beta/H$.
Interestingly, we find that all our calculations result in values for $\beta/H$ which are much larger as compared to the rough estimate of \cref{eq:intro:beta}, namely
\begin{align}
	\beta/H\simeq\mathcal{O}(\num{e4}) \fineq{.}
\end{align}
Since the energy density of the GW background \textit{decreases} \textendash\ depending on the nature of its source \textendash\ linearly or even quadratically with $\beta/H$, and since the spectrum's peak frequency \textit{increases} linearly with $\beta/H$ (see \eg \cite{Caprini2015}), conclusions about the observability of the GW signal change considerably, as we will demonstrate in the main text.

The article is structured as follows.
In \cref{sec:eft}, we briefly introduce the low-energy effective models that we employ throughout the paper.
Those are the Nambu--Jona-Lasinio (NJL) model \cite{Nambu1961a,Nambu1961b} and its Polyakov-loop enhanced variant (PNJL) \cite{Fukushima2003}, as well as the linear sigma model (LSM) \cite{Gell-Mann1960}.
For definiteness, we will consistently focus on the case of QCD-like hidden sectors with \mbox{$\Nc=3$} colors and \mbox{$\Nf=3$} hidden fermion flavors, which we assume to be massless for simplicity.
In \cref{sec:pt,sec:GW} we review important aspects of first-order phase transitions and of the associated GW background spectra, respectively.
\cref{sec:comp} then contains this work's main results.
Namely, we use the aforementioned effective models to explicitly calculate different parameters that are crucial in characterizing the hidden PT.
Furthermore, we determine the corresponding GW signals and evaluate the prospects for their detectability with future observatories by calculating appropriate signal-to-noise ratios.
We present our findings in such a way that they can be applied to hidden or dark sectors over a wide range of inherent scales, \ie, transition temperatures.
Finally, we conclude in \cref{sec:conclusion}.

\section{Low-energy effective models of QCD-like theories}
\label{sec:eft}
\noindent
In the present work we are interested in strongly interacting hidden sectors which we assume to be describable by a theory similar to QCD.
In particular, the hidden analogue of the chiral phase transition ($\chi$PT) will be triggered by non-perturbative effects, requiring a treatment beyond standard perturbation theory.
To that end, we here employ different effective models for the purpose of reliably capturing the phase transition's dynamics.
In order for such an approach to be successful, all of these models necessarily have to share the relevant symmetry properties of the underlying  hidden sector, which we briefly outline in the following.

The matter part of a QCD-like theory with $\Nf$ fermion flavors $q$ transforming in the fundamental representation of a hidden $\suN{\Nc}$ gauge group reads
\begin{align}
	\Lagn_m = \sum_{i=1}^{\Nf} \bar{q}_i ( \I\slashed{D} - m_i ) q_i
	= \sum_{i=1}^{\Nf} \left[ \bar{q}_{Li} \I\slashed{D} q_{Li} + \bar{q}_{Ri} \I\slashed{D} q_{Ri} - m_i (\bar{q}_{Li} q_{Ri}+\hc) \right] \fineq{,}
	\label{eq:model:QCDlagn}
\end{align}
where $i$ is a flavor index.
Color indices ranging from one to \mbox{$\Nc=3$} are suppressed for readability, but are to be understood as being summed over as well.
In the chiral limit of vanishing fermion masses,\mbox{ $m_i=0$}, the above Lagrangian classically exhibits a global chiral symmetry $\mathcal{G}^\prime := \uN{\Nf}_L\times\uN{\Nf}_R$.
Separating the transformations which treat the chiral components $q_L$ and $q_R$ alike (vector subgroup) from those which transform them differently (axial-vector subgroup), as well as using the Lie algebra isomorphism $\uN{N}\cong\suN{N}\times\uN{1}$, the group $\mathcal{G}^\prime$ can be conveniently written as
\begin{align}
	\label{eq:model:Gprime}
	\mathcal{G}^\prime & = \suN{\Nf}_V \times \suN{\Nf}_A \times \uN{1}_V \times \uN{1}_A \fineq{.}
	\intertext{%
It is, however, well known that the axial $\uN{1}$ factor is \textit{explicitly} broken by quantum effects due to the Adler-Bell-Jackiw anomaly \cite{Hooft1976,Hooft1976a}.
The symmetry group of the quantum action is therefore}
	\mathcal{G} & = \suN{\Nf}_V \times \suN{\Nf}_A \times \uN{1}_V \fineq{.}
\end{align}

Nevertheless, the theory's vacuum does not respect the full group $\mathcal{G}$, which is \textit{spontaneously} broken down to \mbox{$\suN{\Nf}_V\times\uN{1}_V$} by a finite expectation value of the fermion condensate, $\chevron{\bar{q}_{i} q_{i}}\neq0$.
However, finite-temperature effects are anticipated to restore the chiral symmetry group $\mathcal{G}$ at sufficiently high temperatures \cite{Dolan1974a,Weinberg1974b,Kirzhnits1976}.
The properties of the associated chiral phase transition depend on both the number $\Nf$ of fermion flavors and their masses, as summarized in the famous Columbia plot \cite{Brown1990}.
Since we are interested in the gravitational wave signatures from a first-order hidden $\chi$PT, we will focus on the scenario of \mbox{$\Nf=3$} massless flavors in the following.
This choice corresponds to the minimal number of flavors needed to obtain a first-order $\chi$PT \cite{Pisarski1984a,Brown1990}.

Apart from the correct symmetry structure, a successful low-energy effective model must, of course, also incorporate all degrees of freedom most relevant to the physics process of interest.
In the case of the chiral phase transition in QCD-like theories, those will primarily be the hidden counterparts of the well-known mesons from real-world QCD, see \eg \cite{Tanabashi2018}.
In particular, the presence of a $\sigma$-like scalar flavor singlet is crucial since it carries the same quantum numbers as the aforementioned fermion condensate $\chevron{\bar{q}_{i} q_{i}}$.
Thus, by acquiring a finite vacuum expectation value (vev) $\sigVEV$ it triggers the spontaneous breakdown of chiral symmetry as indicated above.
In the course of the present work we will additionally take into account the pseudoscalar octet $\vec{\pi}$, the pseudoscalar singlet $\etap$ and the scalar octet $\vec{a}$.
Depending on the concrete model, further degrees of freedom may be taken into consideration.

The remainder of the present section briefly introduces the three effective theories considered throughout this article.
In order to meaningfully compare predictions about gravitational waves as derived from the individual models, we choose their free parameters such that all models produce an identical meson spectrum, \ie the same $\sigma$, $\etap$ and $\vec{a}$ masses, as well as the same pion decay constant $f_\pi$.
Since we exclusively discuss the chiral limit, the pion octet is an exact Nambu-Goldstone boson and thus massless.
The benchmark meson spectra used throughout the paper are listed in \cref{tab:pt:benchmark}.

\begin{table}[t]
	\centering
	\sisetup{round-mode=places,round-precision=0}
	\renewcommand{\arraystretch}{1.4}
	\begin{tabular}{c|S[table-format=3.0]S[table-format=1.0]S[table-format=3.0]S[table-format=3.0]S[table-format=3.0]}
		\toprule
		\multirow{2}{*}{\symhspace{0.5em}{\shortstack{benchmark\\ point}}} & 
		{\multirow{2}{*}{\symhspace{0.3em}{$f_\pi$ [MeV]}}} &
		{\multirow{2}{*}{\symhspace{0.3em}{$m_\pi$ [MeV]}}} &
		{\multirow{2}{*}{\symhspace{0.3em}{$m_\sigma$ [MeV]}}} &
		{\multirow{2}{*}{\symhspace{0.3em}{$m_\etap$ [MeV]}}} &
		{\multirow{2}{*}{\symhspace{0.3em}{$m_a$ [MeV]}}} \\ \\
		\colrule
		A & 72.3 & 0 & 248 & 458 & 491 \\
		B & 90 & 0 & 400 & 671.5 & 696.8 \\
		C & 73.6 & 0 & 291 & 328.353 & 430.8 \\
		D & 107.5 & 0 & 693.5 & 535.251 & 791.7 \\
		\botrule		
	\end{tabular}
	\caption{Benchmark points used throughout the paper. The model parameters which correspond to the shown mass spectra and pion decay constant are compiled in \cref{app:parameters}.}
	\label{tab:pt:benchmark}
\end{table}

\subsection{The Nambu--Jona-Lasinio (NJL) model}
\noindent
The NJL model was originally proposed to describe the nucleus \cite{Nambu1961a,Nambu1961b}.
Nowadays, it is primarily used as a low-energy effective model of strongly coupled theories like QCD.
The NJL model only describes the quarks, as its Lagrangian does not contain any of the mesons which are the physical degrees of freedom at low energies.
As we will explicitly see the Lagrangian contains four- and six-fermion interactions which render the model perturbatively non-renormalizable.
The theory therefore necessarily contains a UV cutoff $\Lambda$.
Following Ref.~\cite{Holthausen2013}, we employ a four-dimensional momentum cutoff scheme and fix \mbox{$\Lambda = \SI{0.93}{GeV}$} throughout the paper.
An extensive review of the NJL model is given in, for example, Ref.~\cite{Klevansky1992}.

The Lagrangian of the three-flavor NJL model in the chiral limit is given by \cite{Holthausen2013}
\begin{equation}
	\Lagn_\tinytext{NJL} = \tr \bar{q} \I \slashed{\partial} q + 2G \tr (\MF^\dagger \MF) + G_D (\det \MF + \hc )
	\qquad\text{with}\qquad
	\MF_{ij} = \bar{q}_j(1-\gamma_5)q_i \fineq{,}
	\label{NJLLagrangian}
\end{equation}
where the $q_i$ are Dirac fields corresponding to the (hidden) quarks, and $\Psi$ is a fermion bi-linear with flavor indices $i,j$ running from $1$ to \mbox{$\Nf = 3$}.
Spinor and color indices are implicit.
As we are only interested in the chiral limit, the above Lagrangian does not contain an explicit quark mass term.
Importantly, the NJL model thus exhibits the global symmetries of massless QCD, see \eg \cite{Hatsuda1994}.
To be more precise, the first two terms are invariant under \mbox{$\mathcal{G}^\prime=\uN{3}_L\times \uN{3}_R$}, whereas the last term is the 't Hooft determinant which mimics the anomalous  $\uN{1}_{A}$ breaking mentioned below \cref{eq:model:Gprime} \cite{Kobayashi1970,Hooft1976a}.
One can explicitly demonstrate that these symmetries are indeed present by realizing that the Dirac spinors $q$ transform under $\uN{1}_V$ and $\uN{1}_A$ as
\begin{alignat}{3}
	\label{eq:model:njl:U1}
	q & \to \E^{\I \vartheta } q \qquad&&\text{and}\qquad\;\; q && \to \E^{\I \vartheta_5 \gamma_5} q \fineq{,} \\
	\intertext{respectively, for \mbox{$\vartheta,\vartheta_5\in\mathds{R}$}. Under $\suN{3}_L$ and $\suN{3}_R$ the chiral components transform as}
	\label{eq:model:njl:SU3}
	q_L & \to U_L q_L &&\text{and}\qquad q_R && \to U_R q_R \fineq{,}
\end{alignat}
respectively, where \mbox{$q = q_L + q_R$}.
The operators $U_L$ and $U_R$ correspond to unitary matrices with determinant equal to one.
The above transformations clearly show the $\mathcal{G}^\prime$ invariance of the kinetic term. 
Establishing the symmetry properties of the other two terms is slightly more involved. 
As a start, note that the bi-linear $\MF$ as defined in \cref{NJLLagrangian} can also be rewritten in terms of the chiral fermion fields, namely \mbox{$\MF_{ij} = \bar{q}_{Rj} q_{Li}$}.
Employing \cref{eq:model:njl:SU3,eq:model:njl:U1}, one then finds that $\MF$ transforms as follows under the (global) symmetries of massless QCD:
\begin{equation}
	\label{eq:NJL:symm}
	\begin{alignedat}{3}
		\MF & \to U_L \MF U_R^\dagger \qquad&&\text{under~~} \suN{3}_L \times \suN{3}_R \fineq{,} \\
		\MF & \to \MF e^{2 \I \vartheta_5 \gamma_5} \ \ &&\text{under~~} \uN{1}_A \fineq{,}\\
		\MF & \to \MF \ \  &&\text{under~~} \uN{1}_V \fineq{.}
	\end{alignedat}
\end{equation}
These transformations show that the second term of the Lagrangian is also invariant under the full group $\mathcal{G}^\prime$.
The last term, the 't Hooft determinant, is trivially $\uN{1}_V$ invariant.
It also exhibits an \mbox{$\suN{3}_L\times \suN{3}_R$} symmetry, because the $\suN{3}$ operators $U_L$ and $U_R$ have unit determinant.
In contrast, \cref{eq:NJL:symm} reveals that $\Psi$ transforms non-trivially under $\uN{1}_A$, which is therefore broken by the 't Hooft determinant.
Thus, the NJL Lagrangian of \cref{NJLLagrangian} has all of the symmetry properties we require for the NJL model to be a suitable effective theory describing the low-energy regime of QCD.

To be able to perform explicit calculations in the NJL model we use the (self-consistent) mean field approximation (MFA) \cite{Kunihiro1984,Hatsuda1985a,Kunihiro1988,Hatsuda1994}. 
Within the MFA one defines the expectation value of the fermion bi-linear $\chevron{\MF}$ as a sum over effective meson fields, namely
\begin{equation}
	-4G \chevron{\MF} = (\sigma + \I \etap) \unitOp + 2(a_a + \I\pi_a) T^a \fineq{.}
	\label{eq:model:njl:meanfield}
\end{equation}
The $\suN{3}$ generators in the fundamental representation $T^a$ are normalized so as to satisfy \mbox{$\tr(T^aT^b)=\delta_{ab}/2$} and fulfill the appropriate $\suN{3}$ algebra relations
\begin{align}
	[T^a,T^b] = \I f_{abc} T^c
	\qquad\text{and}\qquad
	\{T^a,T^b\} = d_{abc} T^c
	\label{eq:model:njl:su3algebra}
\end{align}
with the usual antisymmetric and symmetric $\suN{3}$ structure constants $f_{abc}$ and $d_{abc}$.
Correspondingly, the indices $a$, $b$ and $c$ run from $1$ to \mbox{$\Nf^2-1=8$}.
The meson fields introduced into the theory via \cref{eq:model:njl:meanfield} are to be understood as bosonic auxiliary fields meaning that they do not have tree-level kinetic terms and are thus classically non-propagating.
However, fermionic quantum fluctuations will eventually induce kinetic terms and hence render the fields dynamical.
More details regarding the MFA can be found in \cref{app:NJL}.
In particular, the final form of the mean field Lagrangian in terms of the fermion and effective meson fields is given in \cref{NJLMFA}.

In order to determine the dynamics of the chiral phase transition the model's effective potential is required.
Note that only the $\sigma$ field acquires a finite vev during the $\chi$PT.%
\footnote{The effective \textit{scalar} meson fields are defined in terms of their constituent fermions as \mbox{$\sigma \sim \chevron{\bar{q} \unitOp q}$} and \mbox{$a_a \sim \chevron{\bar{q} T_a q}$}. The isospin symmetry $\suN{3}_V$ present in the Lagrangian is not spontaneously broken, so that all mesons whose definition contains generators with off-diagonal elements, \ie, terms which couple fermions of different flavor, cannot acquire a finite vev. The only mesons whose vev can be non-zero are therefore $\sigma$, $a_3$ and $a_8$. Furthermore,  we assume all quark masses to be \textit{equal}. The vacuum should also respect this symmetry, so that \mbox{$\chevron{\bar{u} u} = \chevron{\bar{d} d} = \chevron{\bar{s} s}$} must hold. Since $\sigma$ is the only meson proportional to the identity matrix, it is the only one which can obtain a finite vev. Note that \textit{pseudoscalars} cannot acquire a non-zero vev because the vacuum is parity even.}
The effective potential will therefore only be a function of the corresponding classical field $\sigCl$, which is defined as the expectation value of $\sigma$ in the presence of some external source. The tree-level potential $\Vzero{}$ can be found in \cref{app:NJL}. 
Apart from the tree-level terms the one-loop vacuum and thermal contributions, $\VCW{}$ and $\VFT{}$, are required and are readily obtained by integrating out the fermions.
Since all mesons are non-propagating at tree-level they do not contribute to the one-loop effective potential: only the fermions contribute.
The finite-temperature effective potential of the NJL model at one loop can thus be written as 
\begin{align}
	\Veff{NJL}(\sigCl,T) = \Vzero{NJL}(\sigCl) + \VCW{NJL}(\sigCl) + \VFT{NJL}(\sigCl,T) \fineq{,}
	\label{eq:model:njl:Veff}
\end{align}
with
\begin{subequations}
\begin{align}
	\Vzero{NJL}(\sigCl) & = \Vtree{NJL}(\sigma=\sigCl,\etap=\pi_a=a_a=0)
	\stackrel{\eqref{eq:mfa:Vtree}}{=} \frac{3}{8G}\sigCl^2 - \frac{G_D}{16 G^3} \sigCl ^3 \fineq{,} \\ 
	\VCW{NJL}(\sigCl) & =-  \frac{3 \Nc}{16 \pi^2} \left[ \Lambda^4 \log \left(1+\frac{M^2}{\Lambda^2}\right) - M^4 \log \left(1+\frac{\Lambda^2}{M^2}\right) +\Lambda^2 M^2\right] \fineq{,} \\ 
	\label{eq:model:njl:VFT}
	\VFT{NJL}(\sigCl, T) & = 6 \Nc \frac{T^4}{\pi^2} \JF(M^2/T^2) \fineq{,} \\
	\label{eq:model:njl:constMass}
	M & \equiv M(\sigCl) = \sigCl - \frac{G_D}{8 G^2} \sigCl^2 \fineq{.}
\end{align}
\end{subequations}
Using the finite-temperature effective potential one can determine the nature of the $\chi$PT by computing the potential's global minimum $\sigMinT(T)$ for each temperature, where \mbox{$\sigVEV=\sigMinT(T=0)$}.
It can be shown that for suitable $G$ and $G_D$ the transition is indeed of first order.

As we will see in \cref{sec:pt}, studying the dynamics of the $\chi$PT necessitates the calculation of the probability that the scalar field which drives the transition, $\sigma$, tunnels from the theory's false vacuum to its true vacuum.
Recall, however, that all meson fields are classically non-propagating in the NJL model.
Since tunneling processes are inherently dynamical, tunneling of the $\sigma$ field can only be described in the current setup, once we compute the $\sigma$ field's kinetic term as it is induced via fermionic quantum fluctuations.
At this stage of the approximation, the auxiliary field $\sigma$ is thus promoted to a propagating quantum field.
In doing so, the crucial quantity is the (finite-temperature) wave-function renormalization, which is defined as
\begin{equation}
	Z_\sigma^{-1} (\sigma) = \left. \frac{\dd \Gamma_{\sigma\sigma}(p^2, \sigma)}{\dd p^2} \right|_{p^2 =0} \fineq{,}
	\label{eq:model:njl:wave_func}
\end{equation}
where $\Gamma_{\sigma\sigma}(p^2, \sigma)$ is the $\sigma$ field's one-loop propagator, \cf \cref{eq:prop1}.
Note that in order to determine $Z_\sigma^{-1}(\sigma)$ this propagator is to be evaluated at the given field value $\sigma$, but not necessarily at $\sigVEV$.
Additionally, the loop integrals in the expression for the propagator quoted in \cref{app:NJL} are generally evaluated at \textit{finite} temperature when calculating $Z_\sigma^{-1}$, and at \textit{zero} temperature to compute the meson pole masses (see below).
The full expression for $Z_\sigma^{-1}$ can be found in \cref{eq:wavefunctren}.

Eventually, we want to compare results about gravitational waves as obtained from the different considered effective models, provided that they predict the same meson mass spectra as well as the same pion decay constant $f_\pi$.
In the NJL model, the masses of the mesons -- in the broken phase -- are derived from the roots of their propagators, which, in turn, are determined from the mean field Lagrangian $\Lagn_\tinytext{MFA}$, see \cref{app:NJL}.
The one-loop meson propagators can be found in \crefrange{eq:prop1}{eq:prop4}.
The pion decay constant on the other hand is given by (see \eg \cite{Klevansky1992,Kohyama2016})
\begin{equation}
	f_\pi ^2  =  4 n_c M_c^2 I_0(\sigVEV)
	\label{eq:model:njl:fPi}
\end{equation}
with the constituent quark mass \mbox{$M_c =  M(\sigVEV)$}, which corresponds to the effective fermion mass defined in \cref{eq:model:njl:constMass} evaluated at the global minimum of the effective potential at \textit{zero} temperature.
The integral $I_0(\sigVEV)$ can be found in \cref{app:integrals}.  

Finally, note that the NJL model only has three free parameters in the chiral limit, namely $\Lambda$, $G$ and $G_D$.
Since we always fix \mbox{$\Lambda = \SI{0.93}{GeV}$} only two free parameters remain.
Therefore, not all of the physical parameters $m_\sigma$, $m_a$, $m_\etap$ and $f_\pi$ are independent of each other.
Using $f_\pi$ and $m_\sigma$ as input to determine $G$ and $G_D$, the other meson masses ($m_a$ and $m_\etap$) become \textit{predictions} of the model.
The pion mass $m_\pi$ is zero since we are working in the chiral limit.

\subsection{The Polyakov loop enhanced NJL model}
\noindent
Real QCD exhibits not only a chiral PT, but also a (de)confinement PT, whose dynamics is governed by the gluons.
Both of the aforementioned phase transitions are known to occur at similar temperatures from lattice QCD, see \eg Ref.~\cite{Fukushima2017}.
Therefore, gluon dynamics might also be relevant to the physics of the chiral phase transition.
The NJL model discussed above does not include any gauge dynamics and is consequently unable to describe confinement.
This issue was remedied in 2004 by Fukushima \cite{Fukushima2003}.
He incorporated gluon dynamics into effective QCD models at finite temperature by adding the Polyakov loop.
This resulted in the so-called Polyakov-enhanced NJL (PNJL) model.
A recent review of the Polyakov loop can be found, for instance, in Ref.~\cite{Fukushima2017}.
The Lagrangian of the PNJL model (again in the mean field approximation) is given by
\begin{equation}
	\Lagn_\tinytext{PNJL}^\tinytext{MFA} = \Lagn_\tinytext{NJL}^\tinytext{MFA} -  V_\text{glue}(L, T) \fineq{,}
\end{equation}
where $\Lagn_\tinytext{NJL}^\tinytext{MFA}$ can be found in \cref{NJLMFA} and $V_\text{glue}$ is the background gluon potential in terms of the deconfinement transition's (pseudo) order parameter $L$, which denotes the expectation value of the traced Polyakov loop \cite{Fukushima2017}.%
\footnote{Our notation is slightly different from the one typically used in the literature (\eg Ref.~\cite{Fukushima2017}), where the expectation value of the traced Polyakov loop is denoted by $\Phi$. Here, we use $L$ to avoid confusion with the meson field $\Phi$.}
Unfortunately, the gluon potential cannot yet be directly calculated from first principles.
Still, the impact of including gauge dynamics can be effectively captured by appropriately parametrizing $V_\text{glue}$.
Data from, for example, lattice QCD can then be used to fix the values of the thus introduced additional free parameters.
A discussion on different parametrizations can be found in Ref.~\cite{Fukushima2008}, where it is also shown that the possible choices for the gluon potential are equivalent for temperatures below two or three times the critical temperature of the deconfinement transition.
The dynamics of the $\chi$PT is hence not expected to depend strongly on the choice of potential.
In this paper we use the gluon potential first introduced in Ref.~\cite{Roessner2006}, namely
\begin{equation}
	T^{-4} V_\text{glue} (L,T) = -\tfrac{1}{2} a(T) L \bar{L} + b(T) \log\bigl[1- 6  L \bar{L} - 3 (L \bar{L})^2 +4(L^3+ \bar{L}^3) \bigr] \fineq{,}
	\label{eq:model:pnjl:Vglue}
\end{equation}
with
\begin{align}
	a(T) = a_0 +a_1 \frac{\Tglue}{T} +a_2 \left(\frac{\Tglue}{T}\right)^2
	\qquad\text{and}\qquad
	b(T) = b_3 \left(\frac{\Tglue}{T}\right)^3 \fineq{.}
	\label{eq:model:pnjl:abFunc}
\end{align}
The potential is chosen such that the value of $L$ is constrained to lie between $0$ and $1$, with the deconfined and confined phases being reached for \mbox{$L \to 1$} and \mbox{$L \to 0$}, respectively.
Furthermore, since we are always working at zero chemical potential, $L$ is real and thus \mbox{$L = \bar{L}$} \cite{Fukushima2017}. 
The critical temperature $\Tglue$ of the QCD deconfinement PT in the pure gauge limit, \ie assuming infinitely heavy quarks, is known to be \mbox{$\Tglue = \SI{270}{MeV}$} from lattice simulations \cite{Schaefer2007}. 
However, when taking into account three mass-degenerate quark degrees of freedom as in our model, $\Tglue$ is reduced to \SI{178}{MeV} \cite{Schaefer2007}. 
Lastly, the dimensionless parameters $a_i$ and $b_3$ in \cref{eq:model:pnjl:abFunc} are determined from lattice QCD calculations as well \cite{Roessner2006}, namely,
\begin{equation}
	\label{eq:PNJLparameters}
	a_0 = 3.51\fineq{,}\qquad
	a_1 = -2.47\fineq{,}\qquad
	a_2 = 15.2\fineq{,}\qquad
	b_3 = -1.75\fineq{.}
\end{equation}

Including the Polyakov loop modifies the thermal part of the NJL model's effective potential from \cref{eq:model:njl:VFT} in two ways.
First, the background gluon potential $V_\text{glue}$ of \cref{eq:model:pnjl:Vglue} is to be added.
Second, the quark coupling to gluons must be taken into account.
In contrast, the vacuum part of the effective potential is unchanged with respect to the NJL model.
In particular, the meson propagators and thus also the meson masses are therefore the same to before.
Combining all of the above, the final expression for the PNJL model's thermal one-loop effective potential is 
\begin{align}
	\Veff{PNJL}(\sigCl, L, T) = \Vzero{PNJL}(\sigCl) + \VCW{PNJL}(\sigCl) + \VFT{PNJL}(\sigCl, L, T)
	\label{eq:model:pnjl:Veff}
\end{align}
with
\begin{subequations}
\begin{align}
	\Vzero{PNJL}(\sigCl) ={}& \frac{3}{8G}\sigCl^2 - \frac{G_D}{16 G^3} \sigCl ^3 \fineq{,} \\ 
	\VCW{PNJL}(\sigCl)={}&-  \frac{3 \Nc}{16 \pi^2} \left[ \Lambda^4 \log \left(1+\frac{M^2}{\Lambda^2}\right) - M^4 \log \left(1+\frac{\Lambda^2}{M^2}\right) +\Lambda^2 M^2\right]  \fineq{,}\\ 
	\VFT{PNJL}(\sigCl, L, T) ={}& - \frac{6 T^4}{\pi^2} \int_0^\infty \!\! \dd x \, x^2 \log \left(1+ \E^{-3\sqrt{x^2+r^2}} + 3 L \E^{-\sqrt{x^2+r^2}} + 3 L \E^{- 2\sqrt{x^2+r^2}}\right) \\
	& + T^4\Bigl(-\tfrac{1}{2} a(T) L^2 + b(T) \log\bigl[1- 6  L ^2 - 3 L^4 +8 L^3 \bigr]\Bigr) \fineq{,}
\end{align}
\end{subequations}
where \mbox{$r\equiv r(\sigCl,T) = M(\sigCl)/T$} with $M$ denoting the effective fermion mass of the NJL model as introduced in \cref{eq:model:njl:constMass}.
The MFA was applied to arrive at the quoted expression for $\VFT{PNJL}$ \cite{Fukushima2017}.

Using the above potential, we find that the deconfinement transition described by the Polyakov loop $L$ is a crossover, while the $\chi$PT with order parameter $\sigMinT$ is of first order for all considered benchmark points.
To proceed we again need to determine how the $\sigma$ field%
\footnote{The wave-function renormalization $Z_\sigma$ is the same in the NJL and the PNJL model.}
tunnels from the theory's false vacuum to its true vacuum.
In doing so, we choose $L$ to minimize the effective potential for any given point $(\sigCl,T)$, \ie \mbox{$L := L_\text{min}(\sigCl, T)$}.
This will reduce $\Veff{PNJL}$ to be only a function of $\sigCl$ and $T$, thus allowing us to use the standard formalism for tunneling in \textit{one} field dimension, which will be outlined in \cref{sec:pt}.
Note that our choice of $L$ is similar to the one made in Ref.~\cite{Kubo2018}.

\subsection{The linear sigma model} \label{sec:model:lsm}
\noindent
Just as the Nambu--Jona-Lasinio (NJL) model, the linear sigma model (LSM), as first introduced by Gell-Mann and Levy \cite{Gell-Mann1960}, is a phenomenological theory to effectively describe certain non-perturbative aspects of the low-energy dynamics of QCD or related strongly coupled theories.
Unlike the NJL model the LSM is a renormalizable, purely scalar theory with the scalar fields being identified with the well-known mesons of QCD.
There exist several variants of the LSM which differ by the global internal symmetry that they are based on.
Since we are interested in a QCD-like theory with \mbox{$\Nf=3$} \textit{massless} fermion flavors, we will here discuss the chiral limit of the \mbox{$\uN{3}\times\uN{3}$} LSM, whose Lagrangian reads as follows:
\begin{align}
	\Lagn_\tinytext{LSM} = \tr \del_\mu\Phi^\dagger\del^\mu\Phi - \Vtree{LSM}(\Phi) \fineq{,}
	\label{eq:model:lsm:Lagn}
\end{align}
where the tree-level potential $V_\text{tree}$ is given by (see \eg \cite{Meurice2017})
\begin{align}
	\label{eq:model:lsm:Vtree}
	\begin{split}
		\Vtree{LSM}(\Phi) ={}& -m^2 \tr(\Phi^\dagger\Phi)
		+ \tfrac{1}{2}(\lambda_\sigma-\lambda_a) \bigl( \tr(\Phi^\dagger\Phi) \bigr)^2
		+ \tfrac{3}{2}\lambda_a \tr\bigl( (\Phi^\dagger\Phi)^2 \bigr) \\
		& + \sqrt{\tfrac{2}{3}} \, c \bigl( \det\Phi + \det\Phi^\dagger \bigr) \fineq{.}
	\end{split}
\end{align}
The complex-valued bosonic field $\Phi$ is a \mbox{$3\times3$} matrix defined as
\begin{align}
	\Phi = \tfrac{1}{\sqrt{6}} (\sigma+\I\etap)\,\unitOp + (a_a + \I\pi_a) \, T^a
	\label{eq:model:lsm:sigma_field}
\end{align}
and thus collects both the scalar meson fields $(\sigma,\vec{a})$ and the pseudoscalar ones $(\etap,\vec{\pi})$.
The $\suN{3}$ generators $T^a$ were introduced below \cref{eq:model:njl:meanfield}, and the index $a$ runs again from 1 to \mbox{$\Nf^2-1=8$}.

Under a general chiral $\mathcal{G}^\prime=\uN{3}_L\times\uN{3}_R$ transformation, the bosonic field $\Phi$ behaves as
\begin{align}
	\Phi \to U_L \Phi U_R^\dagger
	\qquad\text{with}\qquad
	U_{L,R} \in \uN{3}_{L,R} \fineq{,}
	\label{eq:model:lsm:PhiTrafo}
\end{align}
so that $\Phi_{ij}$ has the same quantum numbers  as the fermion bilinear $\bar{q}_{Rj}q_{Li}$, where $q$ are the matter fields introduced in \cref{eq:model:QCDlagn}.
It is straightforward to derive from \cref{eq:model:lsm:sigma_field,eq:model:lsm:PhiTrafo} that $\sigma$ and $\etap$ transform as singlets under the vector subgroup $\uN{3}_V$ of $\mathcal{G}^\prime$, whereas $\vec{a}$ and $\vec{\pi}$ are in its adjoint representation.
Using \cref{eq:model:lsm:PhiTrafo}, one furthermore finds that the kinetic term in \cref{eq:model:lsm:Lagn} and all operators in the first line of \cref{eq:model:lsm:Vtree} are invariant under $\mathcal{G}^\prime$.
In fact, with the exception of the $\lambda_a$-term, all of the aforementioned terms are even symmetric under the larger group $\oN{2\Nf^2}$.
In contrast, the last term in \cref{eq:model:lsm:Vtree} is only invariant under $\mathcal{G}=\suN{3}_V\times\suN{3}_A\times\uN{1}_V$ and thus accounts for anomalous $\uN{1}_A$ breaking \cite{Kobayashi1970,Hooft1976a}.
Spontaneous chiral symmetry breaking $\mathcal{G}\to\suN{3}_V\times\uN{1}_V$ is finally triggered by the scalar singlet $\sigma$ acquiring a finite vacuum expectation value, \ie, $\sigma\to v_\sigma+\sigma$, or in terms of the $\Phi$ field from \cref{eq:model:lsm:sigma_field}
\begin{align*}
	\chevron{\Phi} = \frac{\sigVEV}{\sqrt{6}} \, \unitOp
	\qquad\text{with}\qquad
	\sigVEV = \sqrt{\frac{3}{2}}\,f_\pi \fineq{,}
\end{align*}
where the latter relation is \eg demonstrated in Ref.~\cite{Meurice2017}.

It is well-known that the LSM at finite temperature is plagued by infrared divergences which lead to the breakdown of standard perturbation theory \cite{Dolan1974}.
Many-body resummation schemes are therefore necessary for the purpose of obtaining robust results.
In the following, we will employ the composite operator or 2PI formalism due to Cornwall, Jackiw and Tomboulis (CJT) \cite{Cornwall1974} to implement the so-called Hartree-Fock approximation \cite{hartree_1928,Fock1930}.
Details of the corresponding calculations are relegated to \cref{app:CJT} and can \eg also be found in Refs.~\cite{Petropoulos1999,Lenaghan2000,Lenaghan2000a,Roder2003}.

Here, we only quote the final result for the \mbox{$\uN{3}\times\uN{3}$} LSM finite-temperature effective potential, which reads
\begin{align}
	\Veff{LSM}(\sigCl, T) = \Vzero{LSM}(\sigCl) + \VFT{LSM}(\sigCl,T) \fineq{,}
	\label{eq:model:lsm:Veff}
\end{align}
where $\sigCl$ denotes the expectation value of the $\sigma$ field in the presence of an external source, such that \mbox{$\sigCl\to\sigVEV$} as the source vanishes.
The individual components in \cref{eq:model:lsm:Veff} are given by
\begin{subequations}
	\label{eq:model:lsm:Vcontrib}
	\begin{align}
		\label{eq:model:lsm:V0}
		\Vzero{LSM}(\sigCl) & = \Vtree{LSM}(\Phi=\sigCl\,\unitOp/\sqrt{6}) = -\tfrac{1}{2} m^2 \sigCl^2 - \tfrac{1}{9} c \sigCl^3 + \tfrac{1}{8} \lambda_\sigma \sigCl^4 \fineq{,} \\
		\label{eq:model:lsm:VFT}
		\VFT{LSM}(\sigCl,T) & = \frac{T^4}{2\pi^2} \sum_{i} g_i \Bigl[ \JB(R_i^2) - \tfrac{1}{4}(R_i^2-r_i^2) \IB(R_i^2) \Bigr] \fineq{.}
	\end{align}
\end{subequations}
Note that here we neglect vacuum contributions, which was previously demonstrated to not change the outcome qualitatively \cite{Lenaghan2000,Lenaghan2000a}.
A few further comments regarding \cref{eq:model:lsm:Vcontrib} are in order.
First, the sum in \cref{eq:model:lsm:VFT} runs over all meson species introduced in \cref{eq:model:lsm:sigma_field}, \ie, $i\in\{\sigma,\etap,\pi,a\}$.
The corresponding statistical weights $g_i$ are $g_\sigma=g_\etap=1$ and $g_\pi=g_a=\Nf^2-1=8$.
Moreover, we have defined \mbox{$r_i\equiv r_i(\sigCl,T)=m_i(\sigCl)/T$} and \mbox{$R_i\equiv R_i(\sigCl,T)=M_i(\sigCl,T)/T$}.
Here, the $m_i(\sigCl)$ denote the field-dependent tree-level meson masses \cite{Meurice2017},
\begin{subequations}
	\label{eq:model:lsm:treeLevelMasses}
	\begin{align}
		m^2_\pi(\sigCl) & = -m^2 - \tfrac{1}{3} c \sigCl + \tfrac{1}{2}\lambda_\sigma\sigCl^2 \fineq{,}\\
		m^2_\etap(\sigCl) & = m_\pi^2(\sigCl) + c\sigCl \fineq{,}\\
		m^2_\sigma(\sigCl) & = m_\pi^2(\sigCl) - \tfrac{1}{3}c \sigCl + \lambda_\sigma\sigCl^2 \fineq{,}\\
		m^2_a(\sigCl) & = m_\pi^2(\sigCl) + \tfrac{2}{3} c \sigCl + \lambda_a \sigCl^2 \fineq{,}
	\end{align}
\end{subequations}
while the $M_i(\sigCl,T)$ constitute effective meson masses, which are self-consistently calculated within the CJT formalism for given $T$ and $\sigCl$ as described in \cref{app:CJT}.
Lastly, the thermal integrals $\JB$ and $\IB$ are defined in \cref{eq:integrals:thermal_int} of \cref{app:integrals}.
Formally, the first term in \cref{eq:model:lsm:VFT} reproduces the one-loop term of the conventional perturbative series for the effective potential at finite temperature, see \eg \cite{Quiros1999}.
However, the tree-level masses $m_i$ usually appearing as an argument of the $\JB$ are replaced by the effective masses $M_i$.
The second term goes beyond standard perturbation theory, as well.
In total, it can be shown that the CJT effective potential in the Hartree-Fock approximation accounts for the resummation of all daisy and super-daisy diagrams \cite{Amelino-Camelia1993b}.

Let us finally mention that for the \mbox{$\uN{3}\times\uN{3}$} LSM in the chiral limit, \ie \mbox{$m_\pi=0$}, the number of free Lagrangian parameters $(m^2,\lambda_\sigma,\lambda_a,c)$ equals the number of observable vacuum quantities $(f_\pi,m_\sigma,m_\etap,m_a)$.
Hence, evaluating \cref{eq:model:lsm:treeLevelMasses} at the physical vacuum, \ie for \mbox{$\sigCl=\sqrt{3/2}\,f_\pi$}, yields a system of equations that can be uniquely solved for the former set of parameters given the pion decay constant and a full set of meson masses.

We will use the finite-temperature effective potential of \cref{eq:model:lsm:Veff} for all further calculations related to the linear sigma model.

\section{Chiral phase transition}
\label{sec:pt}
\begin{figure}[t]
	\centering
	\includegraphics[scale=0.9]{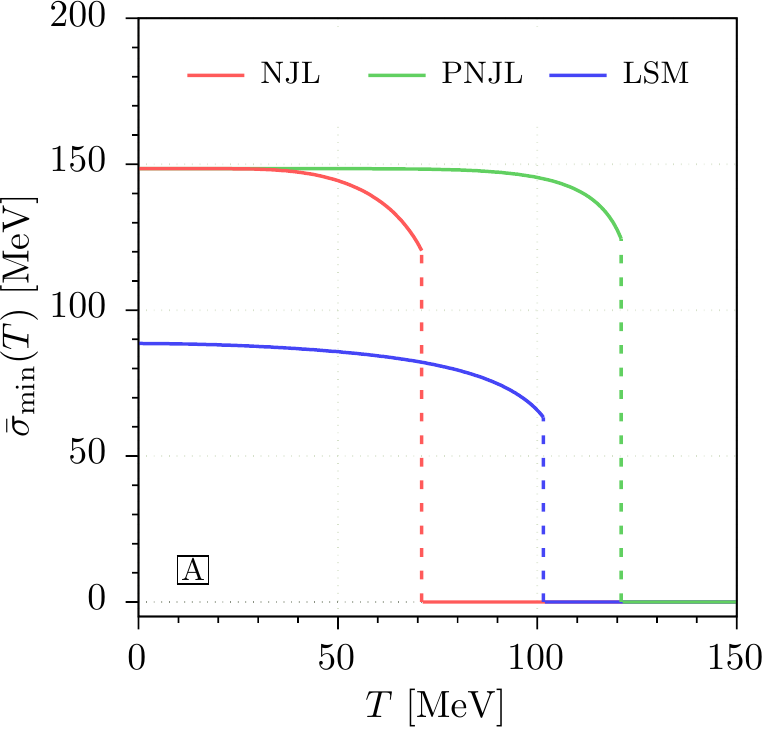}
	\caption{Temperature-dependent global minimum of the effective potential $\sigMinT$ as predicted by the employed low-energy effective models for benchmark point A of \cref{tab:pt:benchmark}. The discontinuity at some critical temperature $T_c$ is characteristic of a first-order phase transition.}
	\label{fig:pt:vsigma}
\end{figure}

\noindent
As we have mentioned in the beginning of \cref{sec:eft}, the chiral phase transition (PT) in QCD-like theories with three (or more) massless fermion flavors is predicted to be of first order on general grounds \cite{Pisarski1984a}.
For the low-energy effective models introduced in the previous section, we can explicitly demonstrate the existence of such a first-order chiral PT by using the appropriate finite-temperature effective potentials in \cref{eq:model:njl:Veff,eq:model:pnjl:Veff,eq:model:lsm:Veff}, respectively.
To that end, we examine how the potential's \textit{global} minimum $\sigMinT$, which, in turn, determines the given theory's true groundstate, changes with temperature.
\cref{fig:pt:vsigma} shows the function $\sigMinT(T)$ as calculated within the investigated low-energy models for one of the benchmark points collected in \cref{tab:pt:benchmark}.
The appearance of two degenerate minima at some critical temperature $T_c$ clearly signals the occurrence of a first-order phase transition.
As an aside, note that the figure also explicitly demonstrates the equivalence of the NJL and the PNJL model at zero temperature.

From a physics point of view, it is well known that a first-order PT proceeds via the nucleation and subsequent growth of bubbles of the true vacuum inside an expanding universe, which is still in the metastable symmetric phase.
Hence, the transition's properties crucially depend on both the nucleation rate $\Gamma$ of the aforementioned bubbles and the Hubble parameter $H$.
In what follows, we will briefly introduce these two quantities in turn.

First, the temperature-dependent bubble nucleation rate $\Gamma$, which is also referred to as the decay rate of the false vacuum, can be calculated as \cite{Linde1981, Hogan1984, Witten1984}
\begin{align}
	\Gamma(T) \simeq T^4 \left( \frac{S_3}{2\pi T} \right)^{\nicefrac{3}{2}} \exp(-S_3/T) \fineq{.}
	\label{eq:pt:Gamma}
\end{align}
In the above equation, the theory's three-dimensional Euclidean action $S_3$ is to be understood as being evaluated for the $\oN{3}$-symmetric bounce or tunneling solution $\sigCl_b(r)$, \ie \mbox{$S_3\equiv S_3[\sigCl_b(r)]$}.
The field configuration $\sigCl_b(r)$, in turn, is determined as the solution of the scalar background field's equation of motion subject to the boundary conditions \mbox{$\sigCl\to0$} as \mbox{$r\to\infty$} and \mbox{$\dd\sigCl/\dd r=0$} at \mbox{$r=0$}, where $r$ is the radial  coordinate of three-dimensional space.
If the $\sigCl$ field represents a \textit{fundamental} degree of freedom, as is \eg the case in the LSM, the theory's Euclidean action can be written as
\begin{align}
	S_3[\sigCl] = \int \!\dd\Omega\dd r \, r^2 \left[ \frac{1}{2}\left(\frac{\dd \sigCl}{\dd r}\right)^{\!2} + V_\text{eff}(\sigCl) \right] \fineq{,}
	\label{eq:pt:actionFundamental}
\end{align}
so that the background field's equation of motion reads
\begin{align}
	\frac{\dd^2 \sigCl}{\dd r^2} + \frac{2}{r} \frac{\dd \sigCl}{\dd r} = \frac{\del V_\text{eff}}{\del \sigCl} \fineq{.}
	\label{eq:pt:eomFundamental}
\end{align}
In contrast, the $\sigCl$ field is \textit{composite} in the (P)NJL model (\ie non-propagating at tree-level).
The Euclidean action therefore has to be slightly modified with respect to \cref{eq:pt:actionFundamental} and is now given by
\begin{align}
	S_3[\sigCl] = \int \!\dd\Omega\dd r \, r^2 \left[ \frac{Z_\sigma^{-1}}{2} \left(\frac{\dd \sigCl}{\dd r}\right)^{\!2} + V_\text{eff}(\sigCl) \right] \fineq{,}
	\label{eq:pt:actionComposite}
\end{align}
where \mbox{$Z_\sigma\equiv Z_\sigma(\sigCl)$} is the wave-function renormalization introduced in \cref{eq:model:njl:wave_func}.
The background field's equation of motion is then found to be
\begin{align}
	\frac{\dd^2 \sigCl}{\dd r^2} + \frac{2}{r} \frac{\dd \sigCl}{\dd r}
	- \frac{1}{2} \frac{\del \log Z_\sigma}{\del \sigCl} \left( \frac{\dd\sigCl}{\dd r} \right)^2 
	= Z_\sigma \frac{\del V_\text{eff}}{\del \sigCl} \fineq{.}
	\label{eq:pt:eomComposite}
\end{align}
Throughout the present work, we solve \cref{eq:pt:eomFundamental,eq:pt:eomComposite} using the \texttt{CosmoTransitions} package \cite{Wainwright2012} and an appropriately customized version thereof, respectively.
The same code is employed to compute the quantity $S_3[\sigCl_b(r)]$ based on \cref{eq:pt:actionFundamental,eq:pt:actionComposite}.

Next, let us discuss the expansion of the universe as it is governed by the temperature-dependent Hubble parameter $H$.
In the absence of significant supercooling, the universe will be radiation dominated during the whole phase transition so that $H(T)$ is given by
\begin{align}
	H^2(T) = \frac{\rho_\text{rad}(T)}{3\MPl^2}
	\qquad\text{with}\qquad
	\rho_\text{rad}(T) = \frac{\pi^2}{30}\geff T^4\fineq{,}
	\label{eq:pt:hubble}
\end{align}
where $\MPl$ is the \textit{reduced} Planck mass, \mbox{$\MPl=\SI{2.435e21}{MeV}$}, while the effective number of relativistic degrees of freedom in the thermal plasma is denoted as $\geff$.

Given the bubble nucleation rate and the Hubble parameter of \cref{eq:pt:Gamma,eq:pt:hubble}, the nucleation temperature $T_n$ is defined as the temperature, for which one bubble per Hubble volume is produced on average, \ie,
\begin{align}
	\int_{T_n}^{T_c} \frac{\dd T}{T} \frac{\Gamma(T)}{H(T)^4} \stackrel{!}{=} 1 \fineq{.}
	\label{eq:pt:Tn}
\end{align}
Typically, the above integral is dominated by temperatures very close to $T_n$.
Hence, the defining condition in \cref{eq:pt:Tn} is approximately equivalent to
\begin{align}
	\Gamma(T_n) \stackrel{!}{=} H(T_n)^4
	\qquad\Longleftrightarrow\qquad
	\left.\frac{S_3(T)}{T}\right|_{T=T_n} = 2 \log \left( \frac{90}{g_*\pi^2}\frac{\MPl^2}{T_n^2} \right) \fineq{,}
	\label{eq:Nuclcond}
\end{align}
where we used \cref{eq:pt:Gamma,eq:pt:hubble} to arrive at the second relation, ignoring the slowly varying factor $(S_3/(2\pi T))^{3/2}$ in the expression for the nucleation rate $\Gamma$.
Notably, the right-hand side of \cref{eq:Nuclcond} depends on the temperature only logarithmically, so that the nucleation condition roughly translates to $S_3/T \approx \mathcal{O}(100)$ for a wide range of temperatures \cite{Hogan1984, Witten1984}.

The gravitational wave spectrum associated with a first-order PT, however, not only depends on the nucleation temperature, but also on various other quantities that characterize the transition's properties.
First, the phase transition's inverse duration $\beta$ compared to the expansion rate of the Universe at the time of the transition is a crucial parameter that can be calculated as%
\footnote{Strictly speaking, Eqs.~\eqref{eq:pt:beta} to \eqref{eq:pt:alpha} are to be evaluated at the \textit{transition} temperature \textendash\ usually denoted as $T_*$ in the literature \textendash\ rather than at the \textit{nucleation} temperature $T_n$.
However, as we will explicitly see in \cref{sec:comp}, the chiral phase transitions under consideration exhibit virtually no supercooling, so that the thermal plasma is practically not reheated after the PT has completed and \mbox{$T_n\simeq T_*$} holds to good approximation.
On a related note, $H$ will throughout the paper refer to the Hubble parameter at the time of the transition unless stated otherwise.
\label{ft:pt:transition}
}
\begin{align}
	\beta = H(T_n)T_n \cdot \left. \frac{\dd (S_3/T)}{\dd T} \right|_{T=T_n} \fineq{.}
	\label{eq:pt:beta}
\end{align}
In the absence of a calculable model \cref{eq:pt:beta} cannot be evaluated without further assumptions on $S_3/T$.
One way is then to follow Refs.~\cite{Hogan1984, Witten1984} and approximate $ \dd S_3/\dd T\gtrsim S_3/T$.
In that case $\beta/H$ can be estimated as
\begin{align}
	\frac{\beta}{H} = \left(\frac{\dd S_3}{\dd T}-\frac{S_3}{T}\right)_{T=T_n}
	\approx \left.\frac{S_3}{T}\right|_{T=T_n}
	\stackrel{\text{Eq.~}\eqref{eq:Nuclcond}}{\approx} 4\log\frac{\MPl}{T_n}
	\approx\mathcal{O}(100) \fineq{,}
	\label{eq:pt:beta:estimate}
\end{align}
where the last relation holds for a wide range of transition temperatures $T_n$.
Since first-principle calculations of $\beta/H$ are lacking for theories exhibiting a chiral phase transition in a strongly coupled hidden sector, the rough estimate in \cref{eq:pt:beta:estimate} was employed for a long time in the corresponding literature, see \eg \cite{Caprini2010,Binetruy2012,Schwaller2015}.
In contrast, the use of effective models in the present work permits us to determine the function $S_3(T)/T$, so that we can evaluate \cref{eq:pt:beta} explicitly.

A second important quantity in the calculation of gravitational wave spectra is usually denoted as $\alpha$ and encodes the energy released during the phase transition.
Correspondingly, it can be regarded as a measure of the PT's strength.
Several definitions of $\alpha$ can be found in the literature, including one in terms of the transition's latent heat normalized to the radiation energy density (see \eg \cite{Ellis:2018mja}),
\begin{subequations}
\label{eq:pt:alpha}
\begin{align}
	\alpha_L & := \frac{1}{\rho_\text{rad}(T_n)} \left( \Delta V_\text{eff}(T_n) - T_n \left. \frac{\del \Delta V_\text{eff}(T)}{\del T} \right|_{T=T_n} \right) \fineq{,}
	\label{eq:pt:alphaL}
	\intertext{%
where \mbox{$\Delta V_{\text{eff}}(T) := V_\text{eff}(0,T) - V_\text{eff}(\sigMinT(T), T)$} with $\sigMinT(T)$ being the temperature-dependent \textit{global} minimum.
Instead of the latent heat one can also use the trace of the energy momentum tensor to define $\alpha$ (see \eg \cite{Espinosa2010,Ellis:2019oqb}),}
	\alpha_T & := \frac{1}{\rho_\text{rad}(T_n)} \left( \Delta V_\text{eff}(T_n) - \frac{1}{4} T_n \left. \frac{\del \Delta V_\text{eff}(T)}{\del T} \right|_{T=T_n} \right) \fineq{.}
	\label{eq:pt:alphaT}
\end{align}
\end{subequations}
Notably, $\alpha_L$ and $\alpha_T$ are equivalent for strong phase transitions with large amounts of supercooling, \ie \mbox{$\Delta V_{\text{eff}} \gg T_n \del \Delta V_\text{eff}(T)/\del T$}.
Since the considered $\chi$PTs will, however, turn out to be weakly first order, a definition for $\alpha$ needs to be chosen.
In the remainder of the paper, we will always set \mbox{$\alpha := \alpha_T$} as required by our use of \cref{eq:GW:kappa} from Ref.~\cite{Espinosa2010}.

\section{Stochastic gravitational wave signal}
\label{sec:GW}
\noindent
The production of gravitational waves (GW) during violent astrophysical processes like black hole mergers is a well-known phenomenon, which is predicted by Einstein's theory of General Relativity \cite{Einstein1918}.
The first GW signals of this kind were recently detected by the LIGO and VIRGO collaborations \cite{Abbott2016, TheLIGOScientific2017, GBM2017}.
Apart from these transient signals due to astrophysical sources, a \textit{stochastic background} of GWs may exist as a result of, for instance, first-order cosmic phase transitions (PT) or an inflationary phase in the early universe \cite{Caprini2018}.
Observing such a background would thus provide a unique opportunity to investigate primordial physics which would otherwise be impossible to study.

As discussed in \cref{sec:pt}, a first-order PT proceeds via the nucleation and subsequent growth of bubbles of the true groundstate.
Gravitational waves are produced when the aforementioned bubbles collide via several different processes, namely, collisions of bubble walls or, equivalently, scalar field shells \cite{Kosowsky1992b}, as well as sound waves \cite{Hindmarsh2014} and magnetohydrodynamic turbulence \cite{Caprini2006} in the thermal plasma.
The relative importance of these contributions is determined by the dynamical properties of the phase transition and hence depends on the underlying particle physics model.
Still, it is possible to broadly distinguish between two different transition scenarios which are usually referred to as \textit{runaway} and \textit{non-runaway}, see \eg \cite{Caprini2015}.

In the former case the friction exerted by the thermal plasma on the nucleated bubbles is too small to slow down their accelerated expansion, which, in turn, is driven by the released latent heat.
Consequently, the bubble wall velocity $v_b$ will continue to increase until it eventually reaches the speed of light $c$. 
In the non-runaway scenario on the other hand, the friction between bubbles and the surrounding plasma is large enough to counteract the bubbles' accelerated expansion.
The bubble wall velocity hence approaches a terminal value, which is bounded by $c$ but may still be relativistic.

In order to determine which of the above cases applies to the QCD-like hidden sector models discussed in the present paper, the following considerations are helpful:
as outlined in \cref{sec:eft}, the chiral phase transition in those models is driven by the $\sigCl$ field.
The latter is expected to have sizable interactions with the theory's remaining degrees of freedom.
The friction exerted by the plasma on the bubbles defined by $\sigCl$ is therefore anticipated to be non-negligible, which already strongly suggests a non-runaway scenario (\ie Case I of Ref.~\cite{Caprini2015}).
Furthermore, we will see in the next section that our calculations predict rather small values for the phase transition strength $\alpha$ defined in \cref{eq:pt:alphaT}, which also indicates a transition of this type.
Accordingly, we will assume in the following that the $\chi$PT proceeds via non-runaway bubbles in all of the considered models and for all of our benchmark points.%
\footnote{Note that, if the $\chi$PT proceeded via runaway bubbles but with the same $\alpha$ and $\beta/H$, the total energy density of the GW background would \textit{decrease} with respect to the non-runaway case.
The reason is that part of the energy which is transformed into bulk motion for non-runaway bubbles will be deposited in the scalar field gradient in a runaway scenario.
The associated contribution to the GW spectrum is, however, suppressed by an additional factor of $(\beta/H)^{-1}$ and is thus irrelevant for \mbox{$\beta\gg H$} \cite{Caprini2015}.
This should be kept in mind in the following.}

In non-runaway scenarios, the fraction of the PT's latent heat that goes into the kinetic energy of the scalar field is negligible, so that the contribution to gravitational wave production from scalar field shell collisions becomes irrelevant \cite{Caprini2015}.
In contrast, and as a result of the large friction between bubbles and the surrounding thermal plasma, a substantial portion of the available energy is converted into bulk motion in the form of sound waves and magnetohydrodynamic turbulence.
In the following, we will briefly discuss both of the aforementioned contributions to GW production in turn, starting with that from sound waves.
The corresponding energy density is \cite{Caprini2015}
\begin{equation}
	\Omega_\text{sw} h^2 = \num{2.65e-6} \; \left(\frac{H}{\beta}\right)  \left(\frac{ \kappa_v \alpha }{1+\alpha}\right)^2 \left(\frac{100}{\geff}\right)^{\nicefrac{1}{3}} v_b \ S_\text{sw}(f) \fineq{,}
	\label{eq:GW:spectrumSW}
\end{equation}
where the spectral shape $S_\text{sw}(f)$ and the spectrum's peak frequency $f_\text{sw}$ are given by 
\begin{equation}
	S_\text{sw}(f) = \frac{f^3}{f_\text{sw}^3}  \left( \frac{7}{4+3 \frac{f^2}{f_\text{sw}^2}} \right)^{\nicefrac{7}{2}}
	\quad\text{and}\quad
	f_\text{sw} = \SI{1.9e-5}{mHz} \; \frac{1}{v_b} \left(\frac{\beta}{H}\right) \left( \frac{T_n}{\SI{100}{MeV}} \right) \left(\frac{\geff}{100}\right)^{\nicefrac{1}{6}} \fineq{,}
	\label{eq:GW:peakSW}
\end{equation}
respectively.
As indicated above, $\alpha$ and $\beta/H$ characterize the strength and inverse duration of the PT (\cf \cref{sec:pt}), while $\geff$ is the effective number of relativistic degrees of freedom in the plasma and $\kappa_v$ denotes the fraction of latent heat converted into bulk motion.
The actual terminal wall velocity $v_b$ of non-runaway bubbles is determined by the details of the underlying particle physics model.
Calculating its specific value is, however, beyond the scope of the present work.
In the following we will instead assume a certain range of different $v_b$ values and investigate how the resulting GW spectra change.
In doing so we will concentrate on highly relativistic bubble wall velocities, \mbox{$v_b\geq\num{0.75}$}, since those lead to the strongest GW signals and are thus most interesting from an observational point of view.
In the considered case of non-runaway bubbles with large wall velocity \mbox{$v_b \lesssim c$} an estimate for the efficiency factor $\kappa_v$ is then given by \cite{Espinosa2010}
\begin{equation}
	\kappa_v \approx \frac{\alpha}{0.73 + 0.083\sqrt{\alpha} +\alpha} \fineq{.}
	\label{eq:GW:kappa}
\end{equation}

That being said, one should keep in mind that the spectrum in \cref{eq:GW:spectrumSW} was derived from simulations assuming that GW production via sound waves is sufficiently long-lasting.
To be more precise, \cref{eq:GW:spectrumSW} is only reliable if \mbox{$\tau_\text{sw}H > 1$} \cite{Hindmarsh2017}, where $\tau_\text{sw}$ is related to the duration of the sound wave period and reads
\begin{align}
	\tau_\text{sw} = (8\pi)^{\nicefrac{1}{3}} \cdot \frac{v_b}{\bar{U}_f \beta}
	\quad\Rightarrow\quad
	\tau_\text{sw}H \propto \left(\frac{\beta}{H}\right)^{-1} \fineq{,}
	\label{eq:GW:tauSW}
\end{align}
with the plasma's root-mean-square four-velocity $\bar{U}_f$.
As we will explicitly see in \cref{sec:comp}, the $\chi$PTs under consideration are throughout predicted to proceed very fast, \mbox{$\beta/H\gtrsim\mathcal{O}(\num{e4})$}, so that the condition \mbox{$\tau_\text{sw}H>1$} is unlikely to be satisfied.
In order to check this explicitly, we follow Refs.~\cite{Espinosa2010,Ellis:2018mja} to first calculate $\bar{U}_f$. Subsequently, we use \cref{eq:GW:tauSW} to determine the product $\tau_\text{sw}H$ assuming \mbox{$v_b=1$}.
Indeed, we find \mbox{$\tau_\text{sw}H = \mathcal{O}(10^{-3})$} across all benchmark points and models.
From a physics point of view, \mbox{$\tau_\text{sw}H<1$} means that sound waves can efficiently source gravitational waves only over a period shorter than a Hubble time, since the transition proceeds too fast.
Correspondingly, \cref{eq:GW:spectrumSW} is expected to \textit{overestimate} the energy density of the produced GWs \cite{Ellis:2018mja}.
The authors of Refs.~\cite{Ellis:2018mja,Ellis:2019oqb} therefore suggested accounting for a potentially shortened sound wave period in the case of fast PTs by multiplying the amplitude of \cref{eq:GW:spectrumSW} by $\tau_\text{sw}H$, namely
\begin{equation}
	\Omega_\text{sw}^\text{fast} h^2 = \tau_\text{sw}H \cdot \Omega_\text{sw} h^2 \fineq{.}
	\label{eq:GW:spectrumSW_fast}
\end{equation}
Note that although the use of this reduction factor is perfectly reasonable from a physics perspective, it is not yet backed up by dedicated numerical simulations, but is rather subject of current research.

The second relevant source of stochastic gravitational waves in the scenario with non-runaway bubbles is the plasma's turbulent motion.
Its contribution to the total GW amplitude is \cite{Caprini2015}
\begin{equation}
	\Omega_\text{tb} h^2 = \num{3.35e-4} \; \left(\frac{H}{\beta}\right)  \left(\frac{ \kappa_\text{tb} \alpha }{1+\alpha}\right)^{\nicefrac{3}{2}} \left(\frac{100}{\geff}\right)^{\nicefrac{1}{3}} v_b \ S_\text{tb}(f) \fineq{,}
	\label{eq:GW:spectrumTB}
\end{equation}
where the spectral shape $S_\text{tb}(f)$ and the spectrum's peak frequency are given by
\begin{equation}
	S_\text{tb}(f) = \frac{f^3}{f_\text{tb}^3} \frac{\left(1+f/f_\text{tb}\right)^{-\nicefrac{11}{3}}}{1+8 \pi  f/\heff} 
	\quad\text{and}\quad
	\label{eq:GW:peakTB}
	f_\text{tb} = \SI{2.7e-5}{mHz} \; \frac{1}{v_b} \left(\frac{\beta}{H}\right) \left( \frac{T_n}{\SI{100}{MeV}} \right) \left(\frac{\geff}{100}\right)^{\nicefrac{1}{6}} \fineq{,}
\end{equation}
with $\heff$ being the Hubble rate at the time of GW production, redshifted to today, namely $\heff = \SI{16.5 e-6}{mHz} \cdot (T_n/\SI{100}{MeV}) \cdot (\geff/100)^{\nicefrac{1}{6}}$.
Typically, one assumes that only a rather small part of the bulk motion is turbulent, namely \mbox{$\kappa_\text{tb}=\epsilon \kappa_v$} with \mbox{$\epsilon\approx\SI{5}{\%}$}, see \eg \cite{Caprini2015,Ellis:2018mja}.
If this is indeed true, the corresponding contribution to GW production will be irrelevant and the stochastic background will arise virtually exclusively from sound waves.
 
However, as it was pointed out \eg in Refs.~\cite{Ellis:2018mja,Ellis:2019oqb}, the above assumption on $\epsilon$ is expected to become invalid in the case of very fast transitions:
After a shortened sound wave period the plasma is anticipated to enter a regime of non-linear dynamics, so that a relatively large amount of the available energy can be transferred to turbulent motion.
The choice \mbox{$\epsilon \approx \SI{5}{\%}$} will therefore probably \textit{underestimate} the associated GW signal if \mbox{$\beta/H\gg1$}.
To account for the described effect, the authors of Ref.~\cite{Ellis:2019oqb} proposed setting \mbox{$\kappa_\text{tb}=\kappa_v$} and multiplying the amplitude of \cref{eq:GW:spectrumTB} by \mbox{$(1-\tau_\text{sw}H)$},%
\begin{equation}
	\Omega_\text{tb}^\text{fast} h^2 = (1-\tau_\text{sw}H) \cdot \bigl. \Omega_\text{tb}h^2 \bigr|_{\kappa_\text{tb}=\kappa_v} \fineq{.}
	\label{eq:GW:spectrumTB_fast}
\end{equation}
Notably, the reduction of the sound wave amplitude quoted in \cref{eq:GW:spectrumSW_fast} together with the above-described enhancement of the GW signal due to turbulence results in both contributions being of almost equal order of magnitude.

As already mentioned before, neither \cref{eq:GW:spectrumSW_fast} nor \cref{eq:GW:spectrumTB_fast} have been derived from numerical simulations.
Rather, they are the result of heuristic arguments and thus may change in the future.
In the following, we will therefore compute the predicted GW spectra both for the conventional case, \ie solely based on \cref{eq:GW:spectrumSW}, and explicitly taking into account the (short) duration of the transition via \cref{eq:GW:spectrumSW_fast,eq:GW:spectrumTB_fast}.

\section{Results}
\label{sec:comp}
\noindent
\begin{table}[t]
	\centering
	\def\myh{0.95em}
	\sisetup{round-mode=places}
	\renewcommand{\arraystretch}{1.4}
	\begin{tabular}{c|l|S[table-format=3.1,round-precision=1]S[table-format=3.1,round-precision=1]S[table-format=2.1,round-precision=1]S[table-format=3.1,round-precision=1]|S[table-format=1.2,round-precision=2]S[table-format=1.1e-1,round-precision=1]}
		\toprule
		\multirow{2}{*}{\symhspace{0.5em}{\shortstack{benchmark\\point}}} &
		\multirow{2}{*}{\symhspace{0.5em}{\shortstack{effective\\model}}} & 
		{\multirow{2}{*}{\symhspace{0.3em}{$T_c$ [MeV]}}} &
		{\multirow{2}{*}{\symhspace{0.3em}{$T_n$ [MeV]}}} &
		{\multirow{2}{*}{\symhspace{0.7em}{$\geff\alpha$}}} &
		{\multirow{2}{*}{\symhspace{0.3em}{$\beta/H$ [\num{e4}]}}} &
		{\multirow{2}{*}{\symhspace{1.5em}{$\gamma$}}} &
		{\multirow{2}{*}{\symhspace{2.2em}{$b$}}} \\
		& & & & & & & \\
		\colrule
		\multirow{3}{*}{A} & \hspace{\myh}NJL & 71.71 & 70.52 & 3.39625 & 1.79   & 1.76 & 1.25e-1 \\
		& \hspace{\myh}PNJL & 121.76 & 121.36 & 1.083 & 9.43   & 1.82 & 5.04e-3 \\
		& \hspace{\myh}LSM & 101.76  & 101.02  & 0.769405 & 4.41  & 1.8641 & 1.8480e-2 \\
		\colrule
		\multirow{3}{*}{B} & \hspace{\myh}NJL & 107.14 & 106.38 & 2.565 & 4.27 & 1.80 & 2.33e-2 \\
		& \hspace{\myh}PNJL & 140.49 & 140.17 & 2.01875 & 13.75 &  1.87 & 2.04e-3 \\
		& \hspace{\myh}LSM & 145.83  & 145.27 & 0.692788 & 8.59 &  1.8944 & 4.5587e-03\\
		\colrule
		\multirow{3}{*}{C} & \hspace{\myh}NJL & 90.84   & 90.58 & 1.22075 & 11.09 & 1.81 & 3.95e-3 \\
		& \hspace{\myh}PNJL & 131.33   & 131.12 & 0.9025 & 45.66  & 1.85 & 2.40e-4 \\
		& \hspace{\myh}LSM & 100.52 & 99.94 & 1.08191 & 5.6649  & 1.8705 & 1.1314e-02 \\
		\colrule
		\multirow{3}{*}{D} & \hspace{\myh}NJL & 180.31  & 180.27 & 0.40375 & 162.59  & 1.92 & 1.39e-5 \\
		& \hspace{\myh}PNJL & 198.27  & 198.25 & 0.3325 & 244.94  &  1.86 & 9.7e-6 \\
		& \hspace{\myh}LSM & 175.25 & 174.51 & 1.21177 & 7.8341 & 1.9089 & 5.0214e-03 \\
		\botrule
	\end{tabular}
	\caption{Parameters characterizing the (hidden) chiral phase transition as predicted by the considered effective models for the benchmark points from \cref{tab:pt:benchmark}. All calculations were performed assuming \mbox{$\geff=\num{47.5}$} which corresponds to \mbox{$\Nf=3$} light hidden fermions and \mbox{$\Nc^2-1=8$} hidden gluons. The contents of the last two columns were obtained by least-square fits of the function $b(1-T/T_c)^{-\gamma}$ to the results of our explicit calculations of $S_3/T$.}
	\label{tab:comp:parameters}
\end{table}%
Having summarized the basics of first-order (chiral) phase transitions and the associated gravitational wave signal in the previous two sections, we will now apply the described formalism to derive predictions for the gravitational wave spectrum within the low-energy effective models introduced in \cref{sec:eft}.
Let us once more stress that using the aforementioned models puts us in the position to explicitly calculate the parameters characterizing the chiral phase transition (\eg $\beta/H$) without having to resort to heuristic arguments or rough estimates.
We will do so in the following, starting with the benchmark points introduced in \cref{tab:pt:benchmark}, for which the transition is expected to occur around the QCD scale.
Building on these results we will, in a second step, also determine the gravitational wave signal originating from hidden chiral phase transitions at (much) higher temperatures.
For the reasons outlined at the end of the last section, the gravitational wave spectra obtained from \cref{eq:GW:spectrumSW} will throughout be compared to those that are based on  \cref{eq:GW:spectrumSW_fast,eq:GW:spectrumTB_fast}.

\boldmath
\subsection{Hidden chiral phase transitions at $\mathcal{O}(100\,\mathrm{MeV})$}
\begin{figure}[t]
	\centering
	\includegraphics[scale=1.07]{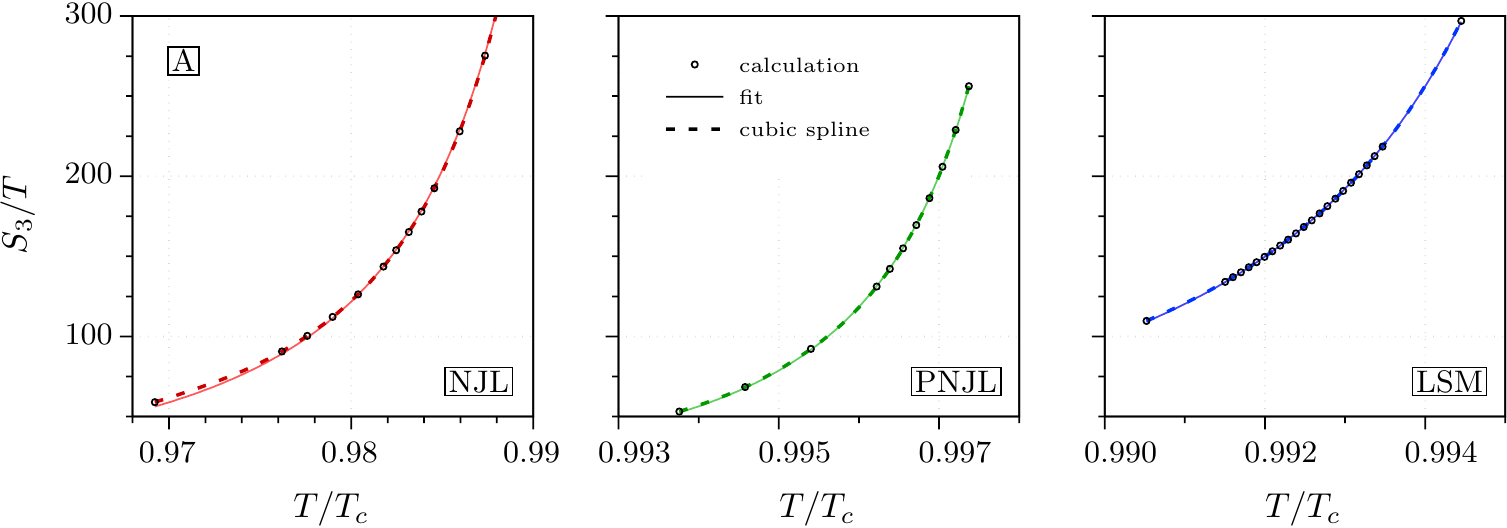}
	\caption{Action functional $S_3/T$ evaluated at the bounce solution as a function of the dimensionless ratio $T/T_c$ for benchmark point A of \cref{tab:pt:benchmark}. For each of the considered effective models, we show the results of our explicit calculations (black circles) alongside the function $b(1-T/T_c)^{-\gamma}$ fitted to the aforementioned data points (solid lines). For comparison, we also plot interpolating cubic splines (dashed lines).}
	\label{fig:results:S3overT}
\end{figure}

\unboldmath
\noindent
Employing the finite-temperature effective potentials of \cref{eq:model:njl:Veff,eq:model:pnjl:Veff,eq:model:lsm:Veff}, we can directly compute the bounce solution $\sigCl_b(r)$ for a given temperature $T$.
\cref{eq:pt:actionFundamental,eq:pt:actionComposite} then allow us to determine $S_3/T$ as a function of the dimensionless ratio $T/T_c$, where the critical temperatures $T_c$ for each benchmark point and model are listed in the first column of \cref{tab:comp:parameters}.
The result of such a calculation is shown in \cref{fig:results:S3overT} for one of our benchmark points.
Interestingly, we observe that $S_3/T$ quite precisely follows a function of the simple form
\begin{align}
	\frac{S_3(T)}{T} \simeq b \left( 1 - \frac{T}{T_c} \right)^{-\gamma} \quad\text{for}\quad T\lesssim T_c \fineq{,}
	\label{eq:results:fit}
\end{align}
which was already anticipated by Hogan in Ref.~\cite{Hogan1984}.
Fitting our data points to the function in \cref{eq:results:fit}, the parameters $\gamma$ and $b$ can be determined, see the last two columns of  \cref{tab:comp:parameters}.
Intriguingly, the exponent $\gamma$ is very similar among all benchmark points \textit{and} effective models.
In contrast, the coefficient $b$ is found to vary by several orders of magnitude.
Using the best-fit values for $\gamma$ and $b$, we plot the function of \cref{eq:results:fit} alongside the data in \cref{fig:results:S3overT} (solid lines).
For comparison, we also show natural cubic splines interpolating between the data points (dashed lines).
All further calculations involving $S_3/T$ will, however, make use of the above described fit, since this is less sensitive to numerical errors.

Given $S_3/T$ as a function of $T$ we can now straightforwardly compute the nucleation temperature $T_n$ by solving \cref{eq:Nuclcond}. 
Most importantly, we always find that \mbox{$T_n \lesssim T_c$} implying that there exists virtually no supercooling, which \textit{a posteriori} justifies our choice of the Hubble parameter in \cref{eq:pt:hubble}, as well as our assumption that the nucleation temperature approximately coincides with the transition temperature (\cf footnote \ref{ft:pt:transition} on page \pageref{ft:pt:transition}).

Next, the phase transition's inverse duration $\beta$ normalized to the expansion rate of the Universe at the time of the transition can be determined from \cref{eq:pt:beta}.
The outcome is again compiled in \cref{tab:comp:parameters}.
Although the precise values for $\beta/H$ vary between different models and benchmark points, we generally find them to be of order \num{e4} or even larger.
Note that these results are in stark contrast to the usual assumption%
\footnote{See also \cref{eq:intro:beta} as well as our discussion in \cref{sec:pt} after \cref{eq:pt:beta}.}
\mbox{$\beta/H \approx \mathcal{O}(100)$}, which is being made in discussions of (hidden) chiral phase transitions throughout the literature,  see \eg \cite{Schwaller2015,Binetruy2012}.%
\footnote{There have been earlier indications that $\beta/H$ can attain much larger values than $\mathcal{O}(100)$, \eg in Refs.~\cite{Tsumura2017,Aoki2017,Bai2019}.}
As we will see in more detail below, this has far-reaching consequences on the observational prospects for the associated gravitational wave signal.
Employing the functional form for $S_3/T$ in \cref{eq:results:fit}, we can even learn why the conventionally used approximation for $\beta/H$ fails.
For that purpose, we compute the derivative
\begin{align}
	\left. \frac{\dd S_3}{\dd T} \right|_{T=T_n} = \left( 1 + \gamma \frac{T_n/T_c}{1-T_n/T_c} \right) \cdot \left. \frac{S_3}{T} \right|_{T=T_n} \fineq{.}
\end{align}
As stated earlier, we always find $T_n$ to be very close to $T_c$ so that the second term in the above expression dominates and $\dd S_3/\dd T$ is much larger than $S_3/T$.
Thus, the central assumption in arriving at $\beta/H\approx S_3/T\approx \mathcal{O}(100)$ via \cref{eq:pt:beta:estimate} is no longer justified.
Alternatively, we can also use \cref{eq:pt:beta,eq:results:fit} to calculate $\beta/H$ directly, giving
\begin{align}
	\frac{\beta}{H} = \gamma \frac{T_n/T_c}{1-T_n/T_c} \cdot \left. \frac{S_3}{T} \right|_{T=T_n} \fineq{,}
	\label{eq:results:beta}
\end{align}
which exhibits the same enhancement with respect to $S_3/T$ for $T_n$ near $T_c$.
Since we generally expect \cref{eq:results:fit} to be valid in the vicinity of the critical temperature, we suspect large $\beta/H\gg S_3/T$ to be a generic feature of models without large supercooling.
As the quantity $S_3/T$ evaluated at $T_n$ varies only very little for a wide range of nucleation temperatures (\cf \cref{eq:Nuclcond}), the exact value of $\beta/H$ then chiefly depends on the ratio $T_n/T_c$, as well as on the exponent $\gamma$.
This behavior is clearly visible in \cref{tab:comp:parameters}, in particular for our benchmark point D.
Here, the (P)NJL model predicts practically degenerate $T_n$ and $T_c$, which leads to huge values for $\beta/H$ of order \num{e6}.

\begin{figure}[t]
	\centering
	\includegraphics[scale=0.92]{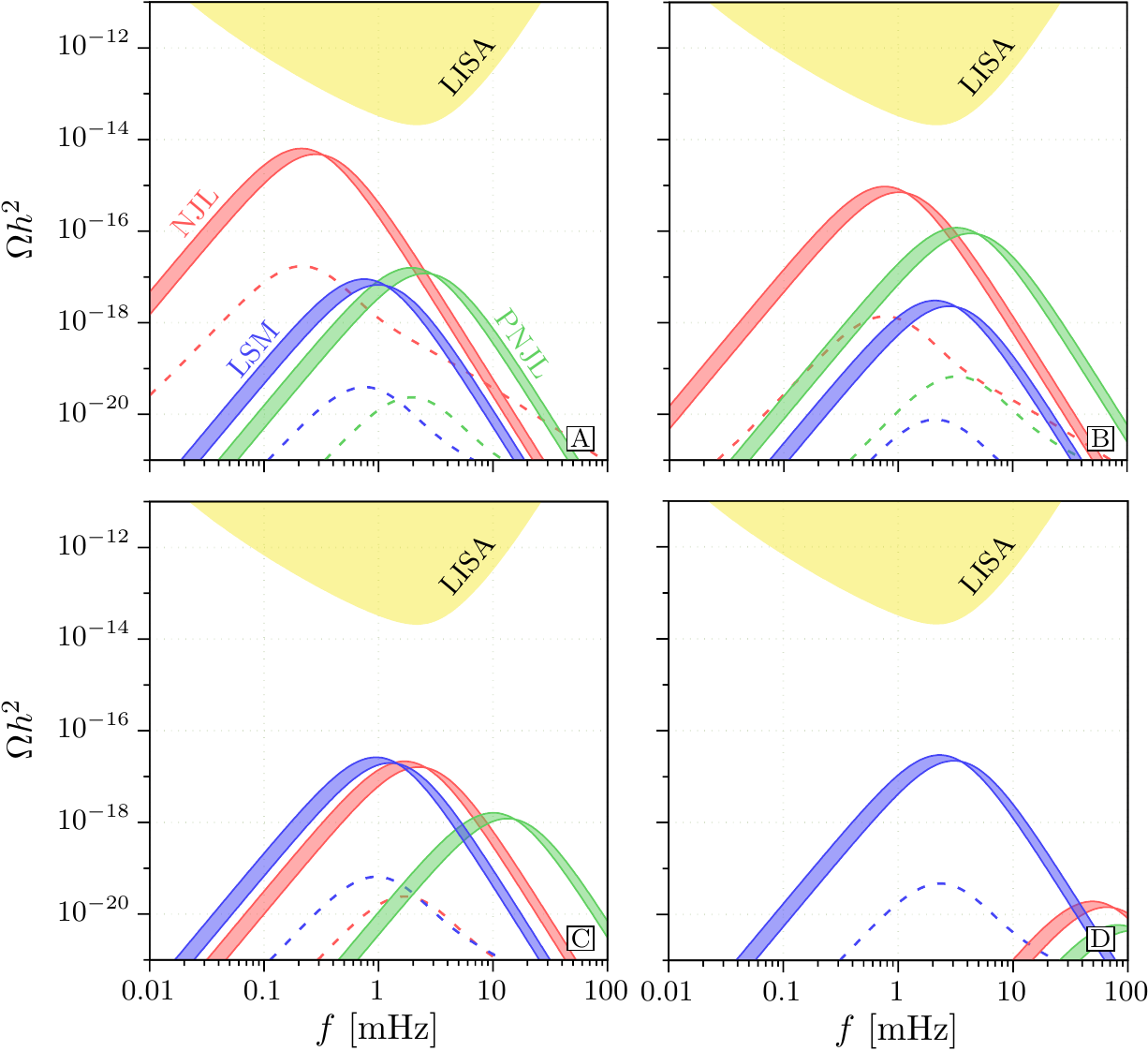}
	\caption{Gravitational wave spectra as predicted for the benchmark points of \cref{tab:pt:benchmark} together with the power-law integrated sensitivity curve \cite{Thrane2013} for the LISA experiment assuming five years of running and a threshold signal-to-noise ratio of $5$. The strain noise power spectral density for LISA was adopted from Ref.~\cite{Yagi2011}. The displayed signal bands were computed from \cref{eq:GW:spectrumSW} for a fixed \mbox{$\geff=\num{47.5}$} by varying the bubble wall velocity between \mbox{$v_b = \num{0.75}$} and \mbox{$v_b = \num{1}$}. The dashed curves show the corresponding spectra obtained from \cref{eq:GW:spectrumSW_fast,eq:GW:spectrumTB_fast} for \mbox{$v_b=1$} and the same $\geff$ as before.}
	\label{fig:results:GW-1e0}
\end{figure}

Lastly, the transition strength $\alpha$ introduced in \cref{sec:pt} is obtained using \cref{eq:pt:alphaT}, see \cref{tab:comp:parameters} for our results.
Let us note that, whereas $T_n$ and $\beta/H$ depend only very mildly on the effective number of relativistic degrees of freedom, $\alpha$ is by definition rather sensitive to the precise value of $\geff$.
By \cref{eq:GW:spectrumSW,eq:GW:kappa} the same is then true for the energy density $\Omega h^2$ of the GW background (but not for its peak frequency, \cf \cref{eq:GW:peakSW}).
To be more precise, it is straightforward to see that \textit{increasing} $\geff$ effectively \textit{decreases} $\Omega h^2$.
For definiteness and unless explicitly stated otherwise, we have fixed \mbox{$\geff=\num{47.5}$} in all calculations, which corresponds to \mbox{$\Nf=3$} light hidden fermions and \mbox{$\Nc^2-1=8$} hidden gluons.
For models with different hidden sectors the presented results must be appropriately adapted.

With the quantities $T_n$, $\beta/H$ and $\alpha$ at hand, and assuming that the chiral phase transition proceeds via non-runaway bubbles with some terminal wall velocity $v_b$, we are now able to compute the predicted gravitational wave spectra $\Omega h^2$ as described in \cref{sec:GW}.
Our findings for the benchmark points from \cref{tab:pt:benchmark} are displayed in \cref{fig:results:GW-1e0}.
Recall that both the position and height of the spectrum's peak are solely determined by the sound wave contribution to $\Omega h^2$.

Importantly, \cref{eq:GW:peakSW} reveals that values of \mbox{$\beta/H\approx\mathcal{O}(\num{e4})$} imply a peak frequency at approximately \SI{1}{mHz} for bubble nucleation at roughly \SI{100}{MeV}.
A first-order chiral phase transition at such temperatures is consequently predicted to produce a stochastic GW background within the LISA frequency band.
At the same time, a large $\beta/H$ suppresses the peak height according to \cref{eq:GW:spectrumSW,eq:GW:spectrumSW_fast,eq:GW:spectrumTB}.
Unfortunately, \cref{fig:results:GW-1e0} seems to suggest that, indeed, the GW signal is too weak to be detected by LISA.
However, comparing the predictions of different effective models for a given benchmark point (\ie a fixed mass spectrum), we find that the parameters characterizing the chiral phase transition as well as the resulting gravitational wave spectrum may vary considerably.
Also the variance across our benchmark points is sizable.
We therefore conclude that our effective model analysis ultimately fails to conclusively answer whether the GW signal can be expected to be observed by LISA.
Consequently, first-principle calculations, like lattice simulations, are required in order to get quantitatively robust predictions.

\subsection{Hidden chiral phase transitions at higher temperatures}
\noindent
\begin{figure}[t]
	\centering
	\includegraphics[scale=0.9]{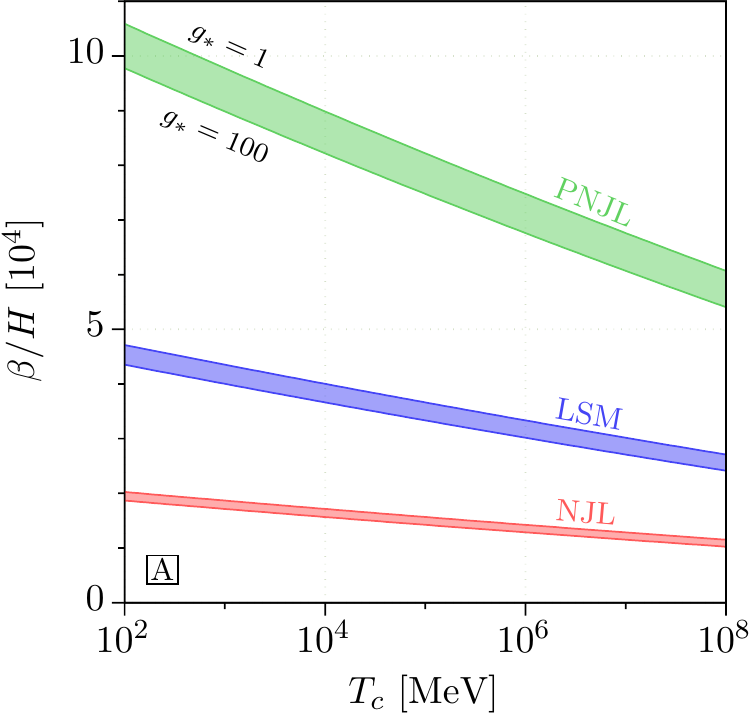}
	\caption{Inverse duration of the hidden chiral phase transition $\beta$ normalized to the Hubble parameter $H$ as a function of the transition's critical temperature for benchmark point A of \cref{tab:pt:benchmark}. The shown bands illustrate the dependence of the result on the effective number of relativistic degrees of freedom $\geff$, which we considered to range from \mbox{$\geff=1$} (upper edge) to \mbox{$\geff=100$} (lower edge).}
	\label{fig:results:beta}
\end{figure}%
In the previous section, we exclusively studied benchmark points with an inherent scale of order \SI{100}{MeV}.
However, as we have argued in the introduction, there are equally well motivated BSM scenarios, where a hidden chiral phase transition is anticipated to occur at higher temperatures.
In order to investigate those cases as well, we will now simply consider scaled-up versions of the benchmark points in \cref{tab:pt:benchmark}, obtained by multiplying all meson masses etc.~by a common dimensionless factor $\xi>1$, \ie,
\begin{align}
	m_i \longrightarrow \xi \cdot m_i
	\fineq{,}\qquad
	f_\pi \longrightarrow \xi \cdot f_\pi
	\qquad\text{and}\qquad
	\Tglue \longrightarrow \xi \cdot \Tglue \fineq{.}
	\label{eq:results:rescaleMasses}
\end{align}
Since it is fully determined by model parameters, the critical temperature $T_c$ will then scale with the same factor.
For the same reason the action functional as a function of $T/T_c$ will remain unaltered, \ie,
\begin{align}
	T_c \longrightarrow \xi \cdot T_c
	\qquad\text{and}\qquad
	\frac{S_3}{T}(T/T_c) \longrightarrow \frac{S_3}{T}(T/T_c) \fineq{.}
\end{align}

In contrast, the nucleation temperature $T_n$ only approximately scales with $\xi$.
An exact scaling is violated by the fact that \cref{eq:Nuclcond} contains the Planck mass as an \textit{absolute} energy scale, which is kept constant.
Accordingly, the ratio $T_n/T_c$ is observed to decrease mildly for growing $\xi$, while it always stays close to one.
A small change in $T_n/T_c$ can, however, have a sizable effect on $\beta/H$ for the rescaled point since $S_3/T$ has a pole at \mbox{$T=T_c$}, close to which its derivative varies rapidly, see \cref{fig:results:S3overT} and \cref{eq:results:beta}.
The resulting scale dependence of $\beta/H$ is displayed in \cref{fig:results:beta}.
Note that $\beta/H$ stays of order \num{e4} for a wide range of critical temperatures in all considered models.
Lastly, in line with its definition in \cref{eq:pt:alphaT} the parameter $\alpha$ is observed to not depend strongly on the precise value of $T_n/T_c$, so that it can be safely assumed as constant in $\xi$ for any fixed choice of $\geff$.

\begin{figure}[t]
	\centering
	\includegraphics[scale=0.92]{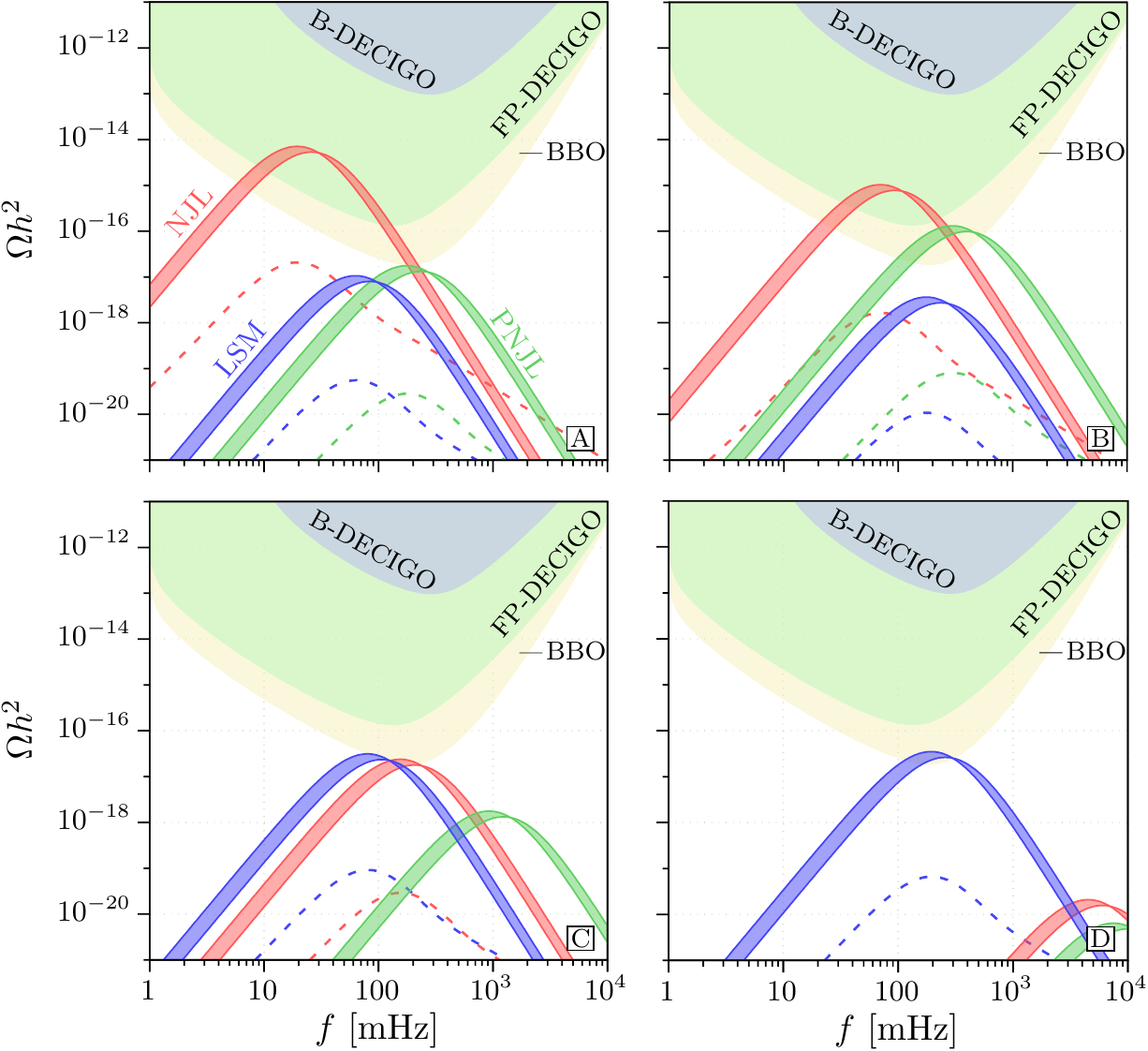}
	\caption{Gravitational wave spectra as predicted for the benchmark points of \cref{tab:pt:benchmark}, but rescaled according to \cref{eq:results:rescaleMasses} with \mbox{$\xi=100$} to obtain transitions at \mbox{$T_n=\mathcal{O}(\SI{10}{GeV})$}. The spectra are shown together with the power-law integrated sensitivity curves \cite{Thrane2013} for various spaceborne GW experiments assuming five years of running and a threshold signal-to-noise ratio of $5$. The strain noise power spectral densities for B-DECIGO, FP-DECIGO and BBO were adopted from Refs.~\cite{Isoyama2018a}, \cite{Yagi2013c} and \cite{Yagi2011}, respectively. The displayed signal bands were computed from \cref{eq:GW:spectrumSW} for a fixed \mbox{$\geff=\num{47.5}$} by varying the bubble wall velocity between \mbox{$v_b = \num{0.75}$} and \mbox{$v_b = \num{1}$}. The dashed curves show the corresponding spectra obtained from \cref{eq:GW:spectrumSW_fast,eq:GW:spectrumTB_fast} for \mbox{$v_b=1$} and the same $\geff$ as before.}
	\label{fig:results:GW-1e2}
\end{figure}

\begin{figure}[t]
	\centering
	\includegraphics[scale=0.92]{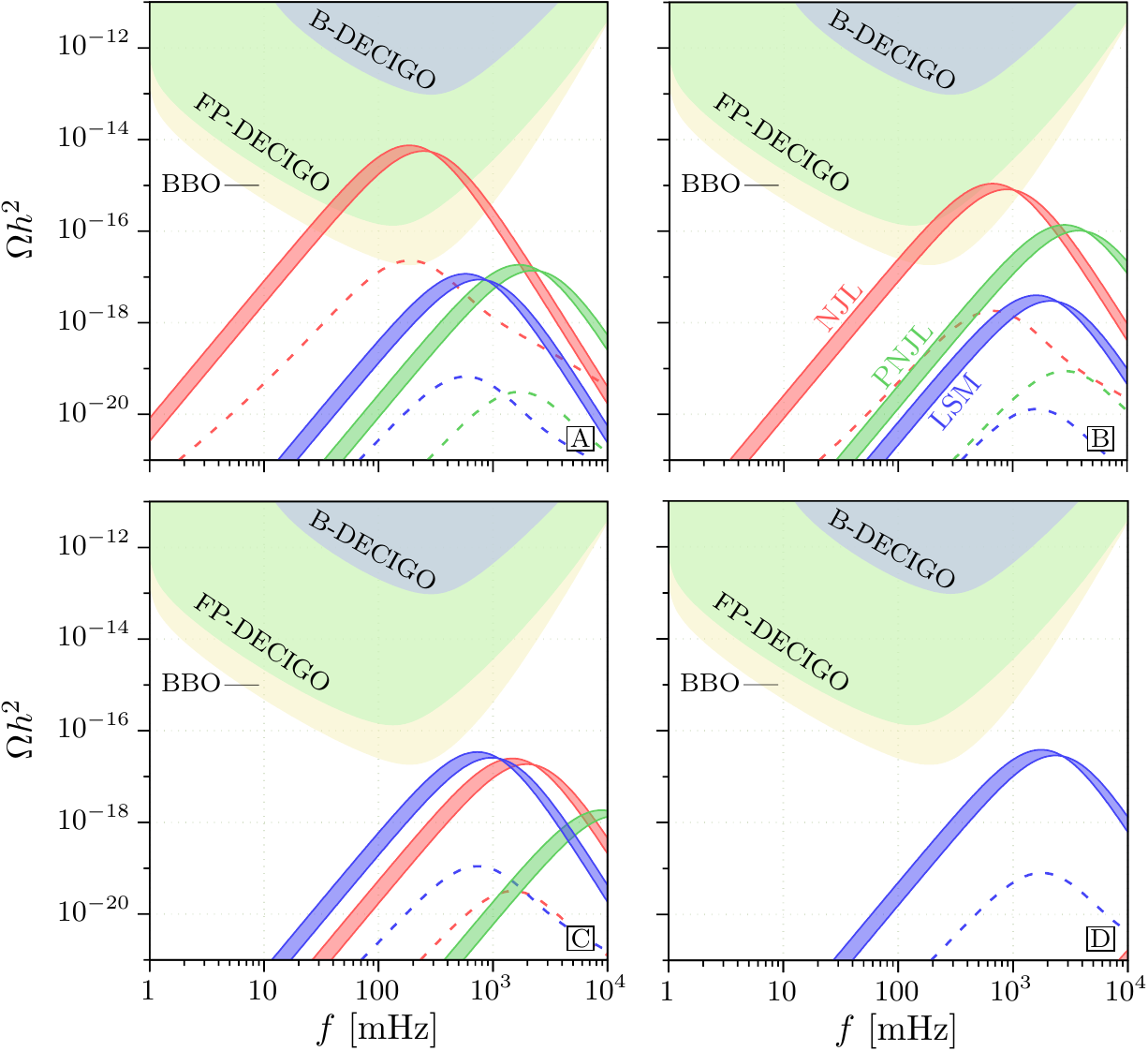}
	\caption{Gravitational wave spectra as predicted for the benchmark points of \cref{tab:pt:benchmark}, but rescaled according to \cref{eq:results:rescaleMasses} with \mbox{$\xi=\num{e3}$} to obtain transitions at \mbox{$T_n=\mathcal{O}(\SI{100}{GeV})$}. The spectra are shown together with the power-law integrated sensitivity curves for various spaceborne GW experiments (\cf caption of \cref{fig:results:GW-1e2} for details). The displayed signal bands were computed from \cref{eq:GW:spectrumSW} for a fixed \mbox{$\geff=\num{47.5}$} by varying the bubble wall velocity between \mbox{$v_b = \num{0.75}$} and \mbox{$v_b = \num{1}$}. The dashed curves show the corresponding spectra obtained from \cref{eq:GW:spectrumSW_fast,eq:GW:spectrumTB_fast} for \mbox{$v_b=1$} and the same $\geff$ as before.}
	\label{fig:results:GW-1e3}
\end{figure}

Following the above discussion and based on the results compiled in \cref{tab:comp:parameters}, we can now compute the GW signal for any given $\xi$.
Importantly, the spectrum's peak frequency is proportional to the nucleation temperature and will thus approximately scale linearly with $\xi$, \cf \cref{eq:GW:peakSW}.
For hidden chiral phase transitions occurring between roughly \SI{1}{GeV} and \SI{10}{TeV} (corresponding to $\xi$ ranging from \num{10} to \num{e5}) the associated GW backgrounds are therefore anticipated to lie within the DECIGO/BBO frequency band.
If we, for the moment, ignore any potential additional suppression of the GW amplitude due to a shortened sound wave period in very fast transitions, our effective model study strongly suggests that the considered experiments are, in fact, sufficiently sensitive for the stochastic background to be observed, \cf the solid bands in  \cref{fig:results:GW-1e2,fig:results:GW-1e3}.
Conversely, the inclusion of the aforementioned effect via the rough estimates in \cref{eq:GW:spectrumSW_fast,eq:GW:spectrumTB_fast} drastically reduces the GW signal, so that a detection becomes less likely.

\begin{figure}[t]
	\centering
	\includegraphics[scale=0.92]{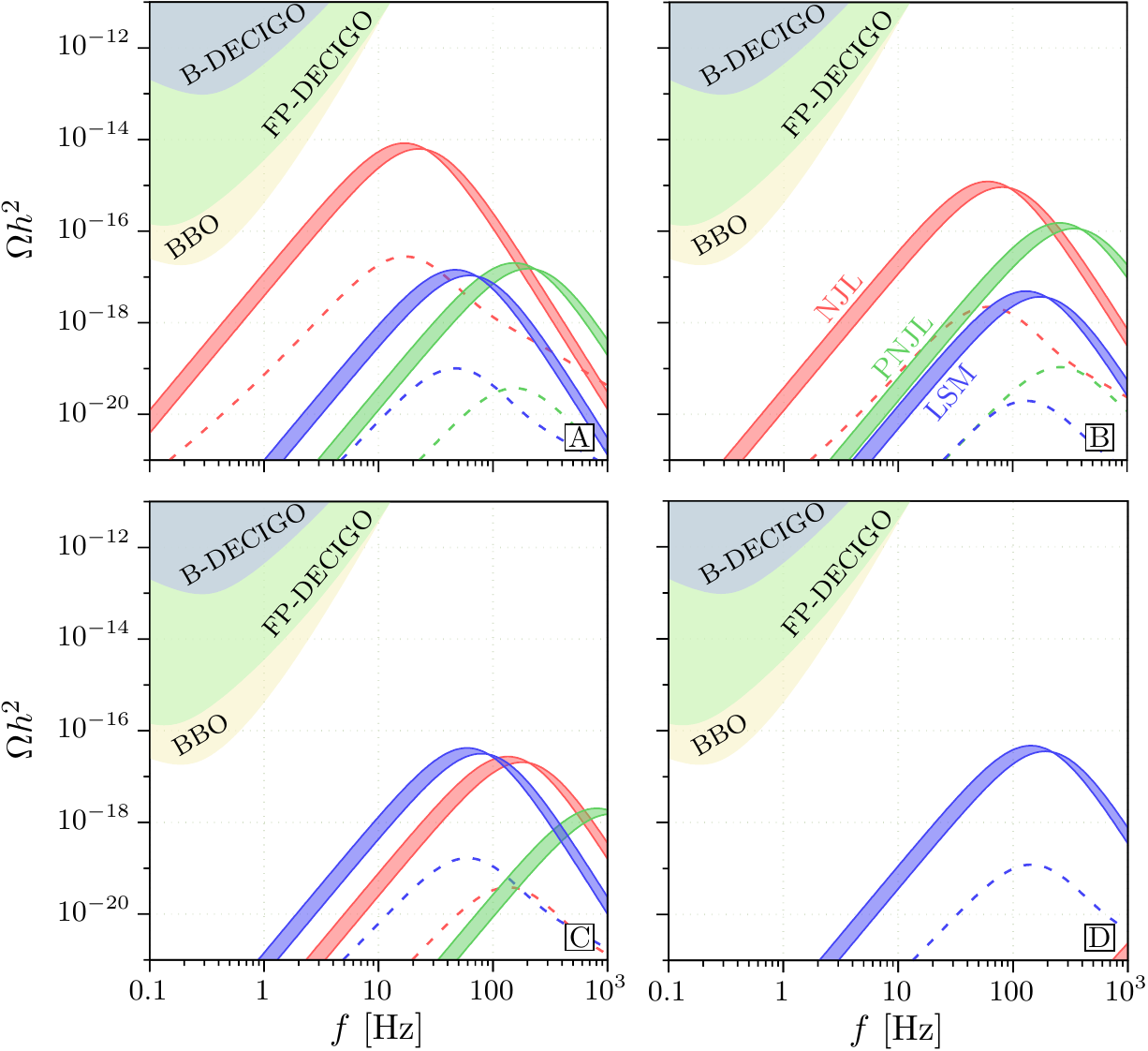}
	\caption{Gravitational wave spectra as predicted for the benchmark points of \cref{tab:pt:benchmark}, but rescaled according to \cref{eq:results:rescaleMasses} with \mbox{$\xi=\num{e5}$} to obtain transitions at \mbox{$T_n=\mathcal{O}(\SI{10}{TeV})$}. The spectra are shown together with the power-law integrated sensitivity curves for various spaceborne GW experiments (\cf caption of \cref{fig:results:GW-1e2} for details). The displayed signal bands were computed from \cref{eq:GW:spectrumSW} for a fixed \mbox{$\geff=\num{47.5}$} by varying the bubble wall velocity between \mbox{$v_b = \num{0.75}$} and \mbox{$v_b = \num{1}$}. The dashed curves show the corresponding spectra obtained from \cref{eq:GW:spectrumSW_fast,eq:GW:spectrumTB_fast} for \mbox{$v_b=1$} and the same $\geff$ as before.
Note that even though the signals lie in the frequency range of ground-based observatories like LIGO, KAGRA or the Einstein Telescope, the sensitivities of those experiments are insufficient for a detection, see \eg \cite{Kuroda2015}}
	\label{fig:results:GW-1e5}
\end{figure}

The above described analysis can, of course, be repeated for larger scaling factors $\xi$ in order to obtain results appropriate to hidden chiral phase transitions at even higher temperatures.
Crucially, we find that for \mbox{$T_n\gtrsim\SI{10}{TeV}$} (or, equivalently, \mbox{$\xi\gtrsim\num{e5}$}) the considered models unanimously predict GW spectra with peak frequencies outside of the sensitivity bands of all planned space-based gravitational wave observatories, \cf \cref{fig:results:GW-1e5}.
A stochastic GW background originating from such high-temperature $\chi$PTs is therefore unlikely to be detectable.

\begin{figure}[t]
	\centering
	\includegraphics[scale=1.0]{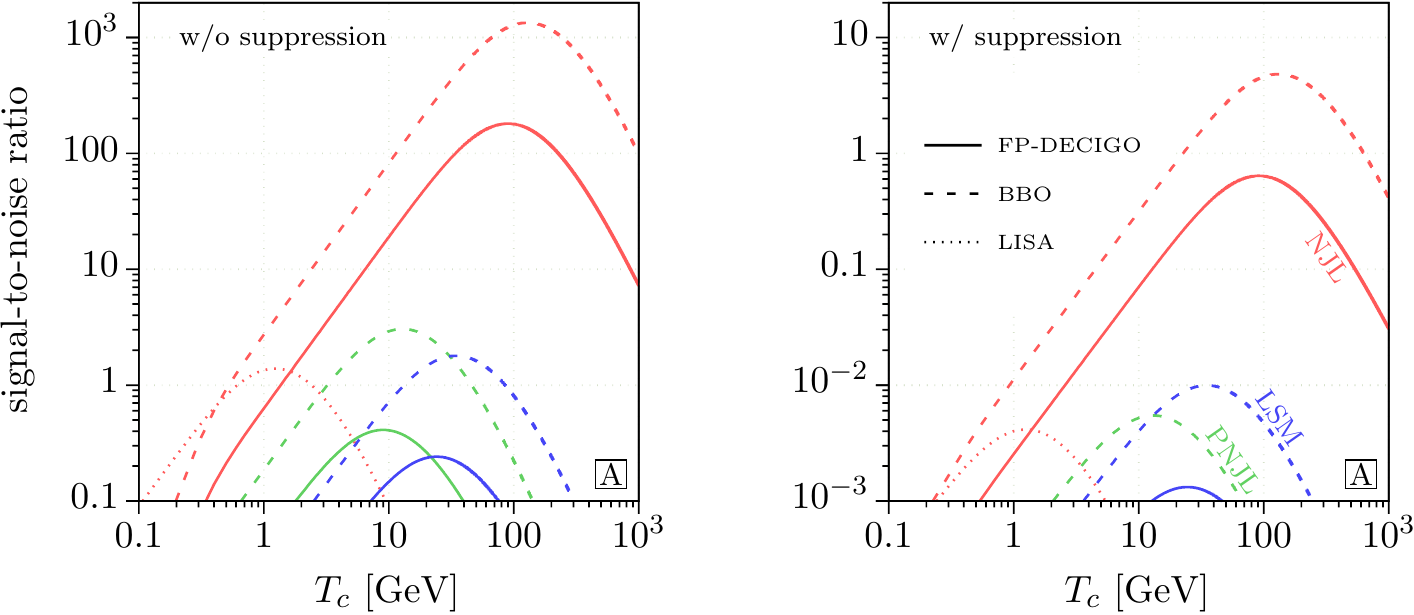}
	\caption{Signal-to-noise ratio as a function of the transition's critical temperature for benchmark point A of \cref{tab:pt:benchmark}. Calculations in which any potential suppression of the GW amplitude due to a shortened sound wave period in very fast transitions is ignored (\textit{left}) are compared to those where such a suppression is incorporated via the heuristic arguments leading to \cref{eq:GW:spectrumSW_fast,eq:GW:spectrumTB_fast} (\textit{right}).}
	\label{fig:results:snr}
\end{figure}

In order to make all of the above qualitative statements slightly more precise, let us finally quantify the discovery potential of the considered experiments by calculating so-called signal-to-noise ratios (SNR), which are defined as (see \eg Ref.~\cite{Thrane2013})
\begin{align}
	\text{SNR} = \sqrt{ 2 t_\text{obs} \int_{f_\text{min}}^{f_\text{max}} \!\!\dd f \left[ \frac{\Omega_\tinytext{GW}(f)\,h^2}{\Omega_\text{noise}(f)\,h^2} \right]^2 } \fineq{.}
	\label{eq:results:snr}
\end{align}
A few comments on the quantities appearing in \cref{eq:results:snr} are in order.
First, $\Omega_\tinytext{GW}(f)h^2$ represents a gravitational wave signal defined by the stochastic background spectra introduced in \cref{sec:GW} and shown in Figs.~\ref{fig:results:GW-1e0} and \ref{fig:results:GW-1e2} to \ref{fig:results:GW-1e5}.
On the other hand, $\Omega_\text{noise}(f)h^2$ denotes a given observatory's effective strain noise power spectral density, expressed as energy density parameter \cite{Moore:2014lga}.
As already mentioned in the caption of Figs.~\ref{fig:results:GW-1e0} and \ref{fig:results:GW-1e2}, we adopt strain noise power spectral densities from Refs.~\cite{Yagi2011,Yagi2013c,Isoyama2018a}.
Frequency integration in \cref{eq:results:snr} is performed over the respective experiment's bandwidth defined by $f_\text{min}$ and $f_\text{max}$.
The duration of an observation in seconds is denoted as $t_\text{obs}$ and will be set to five years in the following.
Let us finally remark that a more complete treatment should also take into account the effect of unresolvable astrophysical foregrounds from black hole, neutron star, and white dwarf mergers on the signal significance, which is, however, beyond the scope of the current analysis.
The resulting signal-to-noise ratios for one of our benchmark points are displayed in \cref{fig:results:snr} and confirm our qualitative discussion from before.

\section{Summary and conclusions}
\label{sec:conclusion}
\noindent
In the present paper, we investigated the prospects for observing a stochastic gravitational wave background associated with a first-order chiral phase transition (PT) in a strongly coupled hidden or dark sector.
For definiteness we studied a new confining $\suN{\Nc}$ gauge force with \mbox{$\Nc=3$} colors acting on \mbox{$\Nf=3$} flavors of massless fermions.
In order to reliably model the strong dynamics that drive the aforementioned transition, we employed various effective theories known to capture non-perturbative aspects of the low-energy regime of real-world QCD: the Nambu--Jona-Lasinio model, its Polyakov-loop enhanced variant and the linear sigma model.

This approach allowed us to explicitly calculate the temperature-dependent action functional $S_3/T$, which is, in turn, crucial to determine key characteristics of the hidden PT like its duration or the nucleation rate of critical bubbles of the true vacuum.
We found that, in the vicinity of its pole at the critical temperature $T_c$, the function $S_3(T)/T$ is of the simple form $b(1-T/T_c)^{-\gamma}$ for some \mbox{$b,\gamma>0$}.
Intriguingly, the exponent $\gamma$ is very similar among all considered benchmark points and effective models, namely \mbox{$\gamma\simeq\numrange[range-phrase=-]{1.8}{1.9}$}.%
\footnote{Interestingly, the found values are very close to the critical exponent $\gamma$ of the mean cluster size in standard percolation theory in three dimensions \cite{Adler1990f,Hellmund2006a,Gracey2015a}. Even though the discussion of a possible connection is beyond the scope of the present article, it might be worthwhile to investigate it in the future.}

As a next step, we then computed the phase transition's nucleation temperature $T_n$, as well as its inverse duration normalized to the Hubble parameter $\beta/H$.
Our results consistently predict the transition to proceed very fast with virtually no supercooling, \mbox{$T_n\lesssim T_c$}.
Crucially, our findings of \mbox{$\beta/H\gtrsim\mathcal{O}(\num{e4})$} are in stark contrast to \mbox{$\beta/H\approx\num{100}$} which is typically assumed in the literature whenever explicit calculations are unavailable.
Lastly, we demonstrated that the observed enhancement of $\beta/H$ by two orders of magnitude has significant impact on the predicted gravitational wave spectrum.
The latter was determined for various transition temperatures appropriate for different well-motivated beyond-the-Standard Model scenarios featuring a new strongly coupled sector.
Our study suggests that a gravitational wave background associated with a hidden chiral PT at temperatures of order \SI{100}{MeV} (\SI{1}{GeV} to \SI{10}{TeV}) may be detected by LISA (DECIGO/BBO).
In contrast, if the transition occurs at temperatures above roughly \SI{10}{TeV} the corresponding gravitational wave signal is unlikely to be observed by any currently proposed experiment.
When looking at these findings, it has to be kept in mind that very fast transitions are generally anticipated to come along with a shortened period of GW production via sound waves and an appropriately suppressed spectrum.
However, there are to date no results from dedicated numerical simulations to incorporate such a suppression in a quantitatively robust way, and thus it is hard to make reliable predictions.

Let us further remark that although all discussed low-energy effective models tend to give qualitatively similar results, the predicted gravitational wave spectra may vary considerably.
This highlights the necessity of true first-principle calculations, like lattice simulations, for the purpose of reaching also quantitative precision.

While the present work focused on the the case of a new $\suN{3}$ gauge sector with three massless fermion flavors, our analysis can, in principle, likewise be applied to different combinations of $(\Nc,\Nf)$.
Those are then, however, not automatically guaranteed to give a first-order PT (see Ref.~\cite{Schwaller2015} for an overview).
We also do not expect our results to change qualitatively if the hidden fermions acquire a (sufficiently small) current mass.
Thus, our study and analogous studies can be employed to investigate the prospects for observing potential gravitational wave signals from a large variety of well-motivated new physics scenarios featuring strongly coupled hidden or dark sectors.

\begin{acknowledgments}
\noindent
We would like to thank Manfred Lindner, Jan M.~Pawlowski, Dirk H.~Rischke, Shinji Takeda, Hong Mao, Masatoshi Yamada, Christian D\"{o}ring, Thomas Hugle, and Vedran Brdar for valuable discussions, as well as Carroll L.~Wainwright for correspondence regarding the \texttt{CosmoTransitions} package. The work of J.K.~is partially supported by the Grant-in-Aid for Scientific Research (C) from the Japan Society for Promotion of Science (Grant No.19K03844).
\end{acknowledgments}

\appendix

\section{Collection of integrals}
\label{app:integrals}
\noindent
In this appendix, we collect integrals used in various expressions throughout the paper.
For instance, the one-loop meson propagators of the (P)NJL model as given in \cref{app:NJL} can be brought into a compact form by defining
\begin{subequations}
	\label{eq:integrals:props}
	\begin{align}
		I_V(\sigVEV) & = \int \! \frac{\dd^4 k}{\I (2 \pi)^4} \, \frac{M_c}{k^2 - M_c^2} \fineq{,} \\
		I_{S}(p^2, \sigVEV) & = \int \! \frac{\dd^4 k}{\I (2 \pi)^4} \, \frac{\tr[(\slashed{k} +\slashed{p} +M_c)(\slashed{k}+M_c)]}{((k+p)^2 - M_c^2)(k^2 - M_c^2)} \fineq{,} \\
		I_{P}(p^2, \sigVEV) & = \int \! \frac{\dd^4 k}{\I (2 \pi)^4} \, \frac{\tr[(\slashed{k} +\slashed{p} +M_c) \gamma_5 (\slashed{k}+M_c) \gamma_5]}{((k+p)^2 - M_c^2)(k^2 - M_c^2) } \fineq{.}
	\end{align}
\end{subequations}
where \mbox{$M_c =  M(\sigVEV)$}, with $M(\sigCl)$ given by \cref{eq:model:njl:constMass}, and $\sigVEV$ denotes the zero-temperature vacuum expectation value of the $\sigma$ field.
All of the above integrals correspond to diagrams which arise from integrating out the fermion fields.
Similarly, the expression for the pion decay constant in the (P)NJL model contains
\begin{equation}
	I_0(\sigVEV) = \int \! \frac{\dd^4 k}{\I (2\pi)^4} \frac{1}{\left(k^2 - M_c^2\right)^2} \fineq{,}
	\label{eq:integrals:fpi}
\end{equation}
Note that the integrals in \cref{eq:integrals:props,eq:integrals:fpi} are divergent and must therefore be regularized.
We employ a hard four-dimensional Euclidean cutoff $\Lambda$ to do so.

Next, the integrals relevant for determining the wave-function renormalization $Z_\sigma$ in the (P)NJL model (\cf \cref{eq:wavefunctren}) are given by
\begin{subequations}
\label{eq:integrals:wavefct}
\begin{align}
\begin{split}
\ell_A(r) ={}& -\frac{1}{4\pi^2}  \int_{0}^{\infty} \!\dd x \;\biggl( \frac{x^2}{\sqrt{x^2 + r^2}^3}  \frac{1}{1 + \exp\sqrt{x^2 + r^2}} \\
& \hspace{7em} + \frac{1}{2} \frac{x^2}{(\sqrt{x^2 + r^2})^2} \frac{1}{1 + \cosh\sqrt{x^2 + r^2}} \biggr) \fineq{,}
\end{split}\\
\begin{split}
\ell_B(r) ={}& \frac{r^2}{16 \pi^2}  \int_{0}^{\infty} \!\dd x \;\biggl( \frac{3 x^2}{\sqrt{x^2 + r^2}^5}  \frac{1}{1 + \exp\sqrt{x^2 + r^2}} +\frac{3 x^2}{2(\sqrt{x^2 + r^2})^4} \frac{1}{1 + \cosh\sqrt{x^2 + r^2}} \\ 
& \hspace{7em} + \frac{x^2}{2(\sqrt{x^2 + r^2})^5} \frac{1}{1 + \cosh\sqrt{x^2 + r^2}} \biggr) \fineq{,}
\end{split}\\
\ell_C(r) ={}& - \frac{r^4}{96 \pi^2}  \int_{0}^{\infty} \!\dd x \;\biggl( \frac{15 x^2}{\sqrt{x^2 + r^2}^7}  \frac{1}{1 + \exp\sqrt{x^2 + r^2}} +\frac{15 x^2}{2(\sqrt{x^2 + r^2})^6} \frac{1}{1 + \cosh\sqrt{x^2 + r^2}} \notag\\ 
& \hspace{7em} + \frac{3 x^2}{(\sqrt{x^2 + r^2})^5} \frac{\tanh(\sqrt{r^2+x^2}/2)}{1 + \cosh\sqrt{x^2 + r^2}} + \frac{x^2}{2(\sqrt{x^2 + r^2})^4} \frac{1}{1 + \cosh\sqrt{x^2 + r^2}} \notag\\ 
& \hspace{7em} - \frac{3 x^2}{2 (\sqrt{x^2 + r^2})^4} \frac{1}{(1 + \cosh\sqrt{x^2 + r^2})^2} \biggr) \fineq{.}
\end{align}
\end{subequations}

Lastly, the bosonic (B) and fermionic (F) thermal integrals needed to compute the one-loop thermal contributions to the effective potentials in \cref{sec:eft} are defined as
\begin{subequations}
	\begin{align}
		\label{eq:integrals:thermal_int:JB}
		\JBF(r^2) & = \pm \int_0^\infty \!\! \dd x \, x^2 \log \left( 1 \mp \E^{-\sqrt{x^2+r^2}} \right) \fineq{,} \\
		\label{eq:integrals:thermal_int:IB}
		\IB(r^2) & = 2\,\frac{\dd \JB(r^2)}{\dd r^2} = \int_0^\infty \!\! \dd x \frac{x^2}{\sqrt{x^2+r^2}} \frac{1}{\E^{\sqrt{x^2+r^2}} - 1} \fineq{.}
	\end{align}
	\label{eq:integrals:thermal_int}
\end{subequations}

\section{Details of the (P)NJL model}
\label{app:NJL}
\noindent
This appendix elaborates on certain aspects of the (P)NJL model and is meant to complement the corresponding discussions in \cref{sec:eft}.
Let us start with a slightly more detailed account of the self-consistent mean field approximation (MFA) \cite{Kunihiro1984,Hatsuda1985a,Kunihiro1988,Hatsuda1994}.
For convenience, we repeat here the Lagrangian for the three-flavor NJL model which is given by
\begin{equation}
	\mathcal{L}_\tinytext{NJL} = \tr \bar{q} \I \slashed{\partial} q + 2G \tr (\MF^\dagger \MF) + G_D (\det \MF + \hc )
	\qquad\text{with}\qquad
	\MF_{ij} = \bar{q}_j(1-\gamma_5)q_i \fineq{.}
\label{NJLLagrangian2}
\end{equation}
Using the MFA, the above Lagrangian can be written in terms of the fermion fields $q_i$ and auxiliary meson fields, \cf also Ref.~\cite{Holthausen2013}.
This is done by going from the perturbative to the Bogoliubov-Valatin (BV) vacuum and using Wick's theorem for operator products normal ordered with respect to the BV vacuum.
Consequently, one can split up the Lagrangian into an \textit{interacting} part $\mathcal{L}_\text{int}$ and a \textit{mean field} term $\mathcal{L}_\tinytext{MFA}$, such that $\mathcal{L}_\tinytext{MFA}$ contains all terms which are at most quadratic in the fermion fields.
One can then rewrite $\mathcal{L}_\tinytext{MFA}$ in terms of $\MF$ and its average $\chevron{\MF}$, which is defined as a sum over mesonic auxiliary fields
\begin{equation}
	-4G \chevron{\MF} = (\sigma + \I \etap) \unitOp + 2(a_a + \I\pi_a) T^a \fineq{,}
\end{equation}
or, equivalently,
\begin{equation}
	\sigma = -\frac{4G}{3} \chevron{\bar{q} q}\fineq{,}\quad
	\pi_a = -4 \I  G \chevron{\bar{q} \gamma_5 T_a q} \fineq{,}\quad
	\etap = - \frac{4 \I G}{3}  \chevron{\bar{q} \gamma_5 q} \fineq{,}\quad
	a_a = - 4 G \chevron{\bar{q} T_a q} \fineq{.}
\end{equation}
Working along the steps outlined above one can thus write the mean field Lagrangian in terms of the effective meson fields $\sigma$, $a_a$, $\pi_a$, $\etap$ and the fermion field $q$.
The NJL Lagrangian in the MFA is given by
\begin{align}
\begin{split}
	\mathcal{L}_\tinytext{NJL}^\tinytext{MFA} ={}&
	\tr \bar{q} (\I \slashed{\partial} - M) q - \I \tr(\bar{q} \gamma_5\pi q)
	- \I \tr(\bar{q} \gamma_5 \etap q) - \tr(\bar{q} a q) \\
	& +\frac{G_D}{8G^2} \Bigl[ (\pi_a \pi_a - a_a a_a -\etap^2 ) \tr (\bar{q} q) - \tr(\bar{q} \pi^2 q) + \tr(\bar{q} a^2 q) \\
& + \tr(\bar{q} \etap\pi q ) + \I \tr(\bar{q} \gamma_5 \sigma \pi q) -2 \I \tr(\bar{q} \gamma_5 \pi a q)  + \I\tr(\bar{q} \gamma_5\etap a q) - \tr(\bar{q} \sigma  a q)  \\
& +\I(3 a_a \pi_a -2 \sigma \etap) \tr(\bar{q} \gamma_5 q) \Bigr] - \Vtree{NJL} \fineq{,}
\end{split}
\label{NJLMFA}
\end{align}
where repeated flavor indices $a$, running from 1 to \mbox{$\Nf^2-1=8$}, are summed over and we have defined $\pi := 2 \pi_a T_a$ and $a := 2 a_a T_a$, which are matrices in flavor space.
The effective fermion mass $M$ was already introduced in \cref{eq:model:njl:constMass} and is given by
\begin{equation}
	M = \sigma - \frac{G_D}{8 G^2} \sigma^2 \fineq{.}
	\label{eq:mfa:M}
\end{equation}
The tree-level potential is
\begin{align}
	\begin{split}
	\Vtree{NJL} ={}&  \frac{1}{8 G} \left( 3\sigma^2 +3\etap^2 + 2 \pi_a \pi_a + 2 a_a a_a \right) - \frac{G_D}{16G^3} \left[ \sigma \left(\sigma^2 + \pi_a \pi_a  - 3\etap^2 - a_a a_a \right) + 5 a_a \pi_a \etap \right] \fineq{.} \\
	\end{split}
	\label{eq:mfa:Vtree}
\end{align}

Next, we briefly discuss the zero-temperature one-loop meson propagators, whose roots define the meson's effective pole masses.
The propagators can be determined based on the mean field Lagrangian of \cref{NJLMFA} by calculating all one-loop 1PI diagrams with two external $\phi$ lines, where $\phi$ stands for one of the meson fields.
The correlator $\I\Gamma_{\phi\phi}$ is then given by the sum of all of these diagrams. 
To find the relevant diagrams one has to keep in mind that only the fermions can run in the loop, because the mesons are represented by auxiliary fields in the NJL model and are thus non-propagating at tree-level.
Furthermore, note that at zero temperature $\sigma$ has a non-zero vev, $\sigVEV$, which implies that diagrams with $\sigVEV$ as an external source also contribute to the meson propagators. 
Taking all these considerations into account, the one-loop meson propagators are computed to be
\begin{subequations}
	\begin{align}
	\label{eq:prop1}
	\Gamma_{\sigma\sigma}(p^2, \sigVEV) & = -\frac{3}{4G} + \frac{3G_D \sigVEV}{8 G^3} -\left(1-\frac{G_D \sigVEV}{4G^2}\right)^2 3 \Nc I_S(p^2, \sigVEV) + \frac{G_D}{G^2}3 \Nc I_V(\sigVEV) \fineq{,} \\
	\label{eq:prop2}
	\Gamma_{\pi\pi}(p^2, \sigVEV) & = -\frac{1}{2G} + \frac{G_D \sigVEV}{8 G^3} +\left(1-\frac{G_D \sigVEV}{8G^2}\right)^2 2 \Nc I_P(p^2, \sigVEV) + \frac{G_D}{G^2}  \Nc I_V(\sigVEV) \fineq{,} \\
	\label{eq:prop3}
	\Gamma_{\etap\etap}(p^2, \sigVEV) & = -\frac{3}{4G} - \frac{3 G_D \sigVEV}{8 G^3} +\left(1+\frac{G_D \sigVEV}{4G^2}\right)^2 3 \Nc I_P(p^2, \sigVEV) - \frac{G_D}{G^2} 3 \Nc I_V(\sigVEV) \fineq{,} \\
	\label{eq:prop4}
	\Gamma_{aa}(p^2, \sigVEV) & = -\frac{1}{2G} - \frac{G_D \sigVEV}{8 G^3} -\left(1+\frac{G_D \sigVEV}{8G^2}\right)^2 2 \Nc I_S(p^2, \sigVEV) - \frac{G_D}{G^2} \Nc I_V(\sigVEV) \fineq{.}
	\end{align}
\end{subequations}
Explicit formulas for $I_V$, $I_S$ and $I_P$ can be found in \cref{app:integrals}. 
Equivalent results are given, for example, in Ref.~\cite{Ametani2015}, where the meson propagators were determined for non-zero quark masses.
Our results agree with theirs when taking the chiral limit of the latter. 

Finally, we provide the full expression for the $\sigma$ field's wave-function renormalization $Z_\sigma$ as obtained within the MFA from its definition in \cref{eq:model:njl:wave_func}:
\begin{equation}
	\label{eq:wavefunctren}
	Z_\sigma^{-1}(\sigma) = -3 \Nc \left(1 - \frac{G_D}{4G^2} \sigma \right)^2 \bigl[ -2 A_0 + 2 B_0 + 8 C_0 - 2 \ell_A(r) + 2 \ell_B(r) + 8 \ell_C(r) \bigr]
\end{equation}
with $r\equiv r(\sigma)=\absVal{M(\sigma)}/T$ and
\begin{align*}
	A_0 & = \frac{1}{16 \pi^2} \left[ \log \left(1 + \frac{\Lambda^2}{M^2}\right) -\frac{\Lambda^2}{\Lambda^2 + M^2} \right] \fineq{,} \\
	B_0 & = -\frac{1}{32 \pi^2} \frac{\Lambda^4}{(M^2 + \Lambda^2)^2} \fineq{,}\qquad
	C_0 = \frac{1}{96 \pi^2} \frac{3 M^2 \Lambda^4 + \Lambda^6}{(M^2 +\Lambda^2)^3} \fineq{,}
\end{align*}
where \mbox{$M\equiv M(\sigma)$} is again the effective quark mass from \cref{eq:mfa:M} and $\Lambda$ denotes the four-dimensional hard momentum cutoff used to regularize the divergent vacuum parts of the occurring loop integrals.
Expressions for the thermal parts $\ell_i$ can be found in \cref{eq:integrals:wavefct} of \cref{app:integrals}.

\section{Basics of the CJT formalism}
\label{app:CJT}
\noindent
The results of the linear sigma model presented in this work's main part were derived within the composite operator formalism due to Cornwall, Jackiw and Tomboulis (CJT) \cite{Cornwall1974}.
Their formalism extends the approach to QFT based on the conventional effective action $\Gamma[\phiCl]$ by employing a generalized functional $\Gamma[\phiCl,\propCl]$, the so-called two-particle irreducible (2PI) action.
Here, $\phiCl$ and $\propCl$ are expectation values%
\footnote{To be more precise, the mentioned quantities are expectation values in the presence of external sources.}
of some quantum field $\Phi$ and of the corresponding two-point function, respectively.
While the original CJT paper \cite{Cornwall1974} only considered the vacuum case \mbox{$T=0$}, the composite operator method was subsequently extended to finite-temperature field theory problems in Refs.~\cite{Amelino-Camelia1993b,Amelino-Camelia1994}, see also \cite{Amelino-Camelia1996}.
It was then employed to calculate the chiral phase transition in different versions of the linear sigma model \eg in Refs.~\cite{Petropoulos1999,Lenaghan2000,Lenaghan2000a,Roder2003}.%
\footnote{Furthermore, a variant of the CJT formalism, the so-called symmetry-improved CJT method \cite{Pilaftsis2013}, was applied to the $\oN{4}$ linear sigma model in Ref.~\cite{Mao2014}.}
In the present appendix, we will only give a brief overview of the CJT formalism's very basics.
For more details we refer the interested reader to the given references.

As mentioned above, the 2PI effective action is a functional of the generally spacetime-dependent expectation values $\phiCl(x)$ and $\propCl(x,y)$, with a given theory's physical solutions satisfying the following stationarity conditions:
\begin{align}
	\frac{\delta \Gamma(\phiCl,\propCl)}{\delta \phiCl(x)} = 0
	\qquad\text{and}\qquad
	\frac{\delta \Gamma(\phiCl,\propCl)}{\delta \propCl(x,y)} = 0 \fineq{.}
	\label{eq:cjt:stationarity}
\end{align}
In situations with translational invariance, however, $\phiCl$ becomes constant and $\propCl$ can be taken as a function of $(x-y)$ only.
All relevant information is then encoded in the 2PI effective potential $\VCJT{}[\phiCl,\propCl]$.
In the chiral limit of the \mbox{$\uN{3}\times\uN{3}$} linear sigma model (LSM) discussed in \cref{sec:model:lsm}, where mixing between the different meson species is absent, the aforementioned potential can be written as \cite{Lenaghan2000a,Roder2003}
\begin{align}
	\VCJT{LSM}[\sigCl, \propCl(k)] = V_0(\sigCl) + \tfrac{1}{2} \sum_i g_i \left( \int_k \log \propCl_i^{-1}(k) + \tfrac{1}{2} \int_k \left[ \propTree_i^{-1}(k;\sigCl) \propCl_i(k) - 1 \right] \right) + V_2[\sigCl,\propCl(k)] \fineq{.}
	\label{eq:cjt:VCJT}
\end{align}
Here, the sum runs over all meson species and the corresponding multiplicities $g_i$ are given by \mbox{$g_\sigma=g_\etap=1$} and \mbox{$g_\pi=g_a=8$} for \mbox{$\Nf=3$}.
Finite-temperature momentum space integration is abbreviated as
\begin{align}
\int_k f(k) := T\sum_{n=-\infty}^\infty \int \! \frac{\dd^3 \vec{k}}{(2\pi)^3} \; f(\omega_n, \vec{k})
\end{align}
with \mbox{$\omega_n=2\pi nT$} being the bosonic Matsubara frequencies.
Furthermore, $V_0$ is the model's tree-level potential given in \cref{eq:model:lsm:V0}, while the $\propTree_i$ represent the tree-level meson propagators, \ie
\begin{align}
	\propTree_i^{-1}(k;\sigCl) = k^2 + m_i^2(\sigCl) \fineq{,}
\end{align}
with the field-dependent tree-level meson masses $m_i$ of \cref{eq:model:lsm:treeLevelMasses}.
Finally, the quantity $V_2$ is computed as the sum of all \textit{two}-particle irreducible vacuum graphs with each line corresponding to a full propagator $\propCl$, as well as generally $\sigCl$-dependent vertices.

Importantly, the conventional effective potential $V_\text{eff}$ as a function of the classical field $\sigCl$ is then obtained from \cref{eq:cjt:VCJT} as
\begin{align}
	\Veff{LSM}(\sigCl) := \VCJT{LSM}[\sigCl,\propPhys(k;\sigCl)] \fineq{,}
	\label{eq:cjt:Veff}
\end{align}
where the full, field-dependent propagators $\propPhys_i$ are determined via appropriate gap equations, which derive from the second stationarity condition in \cref{eq:cjt:stationarity}, and read
\begin{align}
	\left. \frac{\delta \VCJT{LSM}[\sigCl, \propCl(k)]}{\delta \propCl_i(k)} \right|_{\propCl(k) = \propPhys(k;\sigCl)} = 0 \fineq{.}
\end{align}
Using the CJT effective potential from \cref{eq:cjt:VCJT}, the above gap equations can be equivalently written as
\begin{align}
	\propPhys_i^{-1}(k;\sigCl) = \propTree_i^{-1}(k;\sigCl) + \hat{\Pi}[\sigCl,\propPhys(k;\sigCl)]
	\quad\text{with}\quad
	\hat{\Pi}[\sigCl,\propPhys(k;\sigCl)] := 2\left. \frac{\delta V_2[\sigCl, \propCl(k)]}{\delta \propCl_i(k)} \right|_{\propCl(k) = \propPhys(k;\sigCl)}
	\label{eq:cjt:selfenergy}
\end{align}
being the scalar self-energy.
Note that the well-known form of the one-loop one-particle irreducible effective potential $V_\text{eff}^\mathsmaller{(1)}$ will be exactly reproduced from \cref{eq:cjt:Veff}, if \mbox{$V_2\equiv0$} is assumed.
Consequently, any non-trivial improvement over $V_\text{eff}^\mathsmaller{(1)}$ requires a finite $V_2$.
On the other hand, an exact determination of $V_2$ is in general not feasible since it involves the calculation of infinitely many diagrams.
In practice, one must therefore resort to truncations of $V_2$.

\begin{figure}[t]
	\centering
	\subfloat[\label{fig:CJT:2loop:double}]{\includegraphics[scale=0.36]{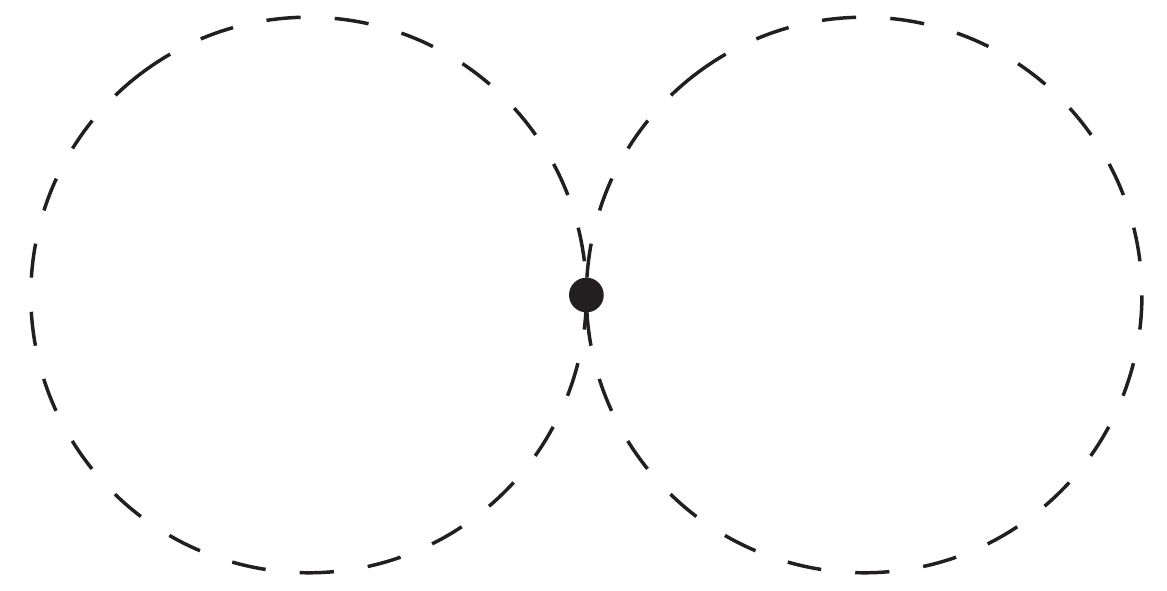}}%
	\hspace{1em}%
	\subfloat[\label{fig:CJT:2loop:sunset}]{\includegraphics[scale=0.36]{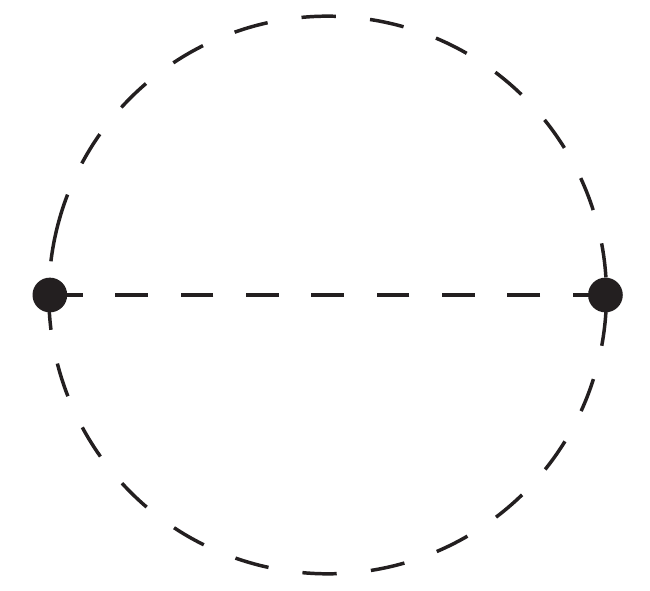}}%
	\hspace{0.7em}%
	\raisebox{2.8em}{$\boldsymbol{\xrightarrow{\hspace*{2em}}}$}%
	\hspace{0.7em}%
	\subfloat[\label{fig:CJT:2loop:tadpole}]{\includegraphics[scale=0.36]{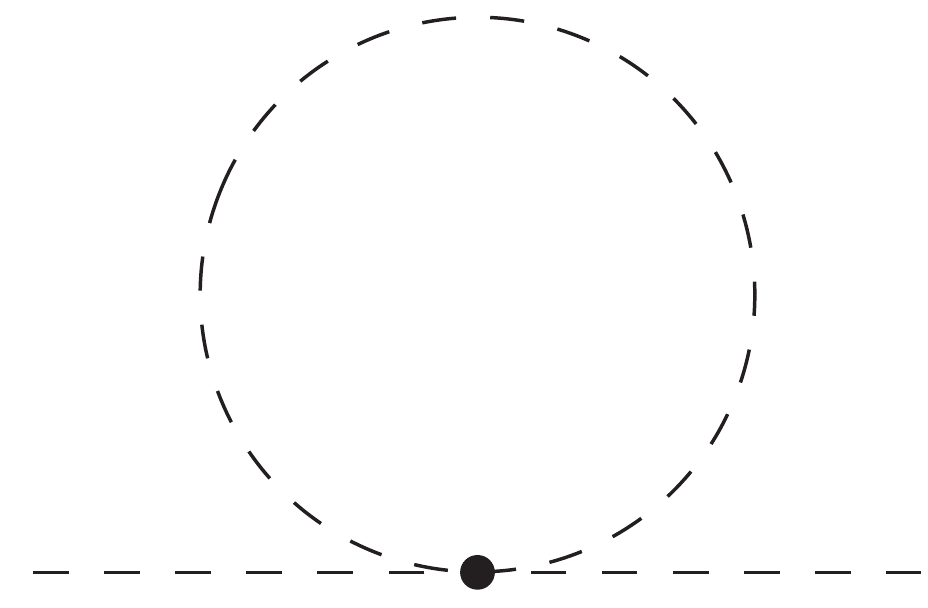}}%
	\hspace{1em}%
	\subfloat[\label{fig:CJT:2loop:normal}]{\includegraphics[scale=0.36]{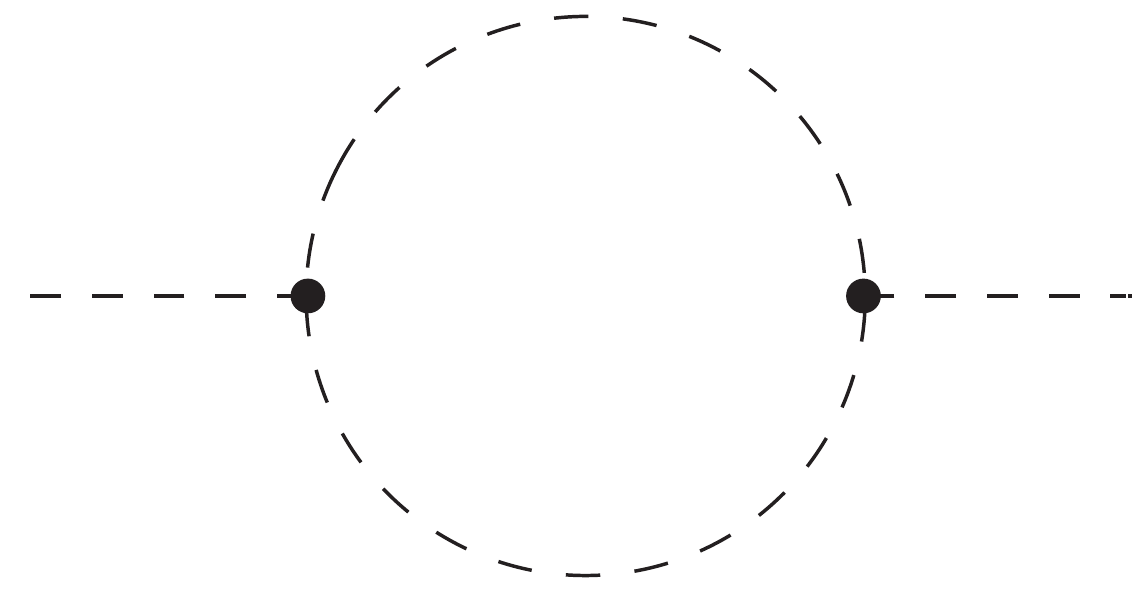}}%
	\caption{Possible topologies of two-loop diagrams contributing to $V_2$ (graphs (a) and (b)), as well as the associated one-loop contributions to the scalar self-energy (graphs (c) and (d)).
All dashed lines correspond to full, field-dependent mesonic propagators $\propCl_i$, with $i$ being either $\sigma$, $a$, $\etap$ or $\pi$. 
Vertices are obtained from the shifted LSM tree-level potential \mbox{$\Vtree{LSM}(\PhiCl+\Phi)$}, where $\Vtree{LSM}$ and the collective meson field $\Phi$ are given in \cref{eq:model:lsm:Vtree,eq:model:lsm:sigma_field}, respectively, while \mbox{$\PhiCl:=\sigCl\unitOp/\sqrt{6}$}.
In the Hartree-Fock approximation used in the present paper, only diagrams of the double-bubble topology (a) are taken into account in calculating $V_2$, so that only tadpole graphs (c) contribute to the self-energy.}
	\label{fig:CJT:2loop}
\end{figure}

Following Refs.~\cite{Petropoulos1999,Lenaghan2000,Lenaghan2000a,Roder2003}, we use the so-called Hartree-Fock approximation \cite{hartree_1928,Fock1930} throughout the present work.
At two-loop level, this corresponds to only retaining contributions from double-bubble graphs (\cf \cref{fig:CJT:2loop:double}) in computing $V_2$ \cite{Cornwall1974}.
In contrast, sunset diagrams as the one displayed in \cref{fig:CJT:2loop:sunset} are consistently ignored.
According to its definition in \cref{eq:cjt:selfenergy} the scalar self-energy thus exclusively obtains contributions from graphs of the tadpole topology shown in \cref{fig:CJT:2loop:tadpole}.

In order to solve the system of gap equations in \cref{eq:cjt:selfenergy} it is now generally convenient to make an \textit{ansatz} for the full field-dependent propagator, namely
\begin{align}
	\propPhys_i^{-1}(k;\sigCl) = k^2+M_i^2(\sigCl,k) \fineq{.}
	\label{eq:cjt:ansatz}
\end{align}
In the Hartree-Fock approximation, the effective meson masses $M_i$ are the tree-level masses $m_i$ dressed by the previously mentioned tadpole contributions and can therefore be assumed to be momentum independent, \ie $M_i\equiv M_i(\sigCl)$, see \eg \cite{Amelino-Camelia1993b}.
\cref{eq:cjt:selfenergy} consequently reduces from a set of integral equations in $\hat{G}_i(k;\sigCl)$ to a system of fixed-point equations in the effective masses $M_i(\sigCl)$.
Explicit calculations eventually yield \cite{Lenaghan2000a,Roder2003}
\begin{subequations}
	\begin{align}
		M^2_\sigma(\sigCl) & = m^2_\sigma(\sigCl) + \frac{T^2}{4\pi^2} \Bigl[ 3\lambda_\sigma \IB(R^2_\sigma) + 8(\lambda_\sigma+2\lambda_a)\IB(R^2_a) + \lambda_\sigma\IB(R^2_\etap) + 8 
\lambda_\sigma\IB(R^2_\pi) \Bigr] \fineq{,} \\
		\begin{split}
			M^2_a(\sigCl) & = m^2_a(\sigCl) + \frac{T^2}{4\pi^2} \Bigl[ (\lambda_\sigma+2\lambda_a) \IB(R^2_\sigma) + 5(2\lambda_\sigma+\lambda_a)\IB(R^2_a) \\
			& \hspace{7.5em} + \lambda_\sigma\IB(R^2_\etap) + (8\lambda_\sigma+9\lambda_a) \IB(R^2_\pi) \Bigr] \fineq{,}
		\end{split}\\
		M^2_\etap(\sigCl) & = m^2_\etap(\sigCl) + \frac{T^2}{4\pi^2} \Bigl[ 3\lambda_\sigma \IB(R^2_\etap) + 8(\lambda_\sigma+2\lambda_a)\IB(R^2_\pi) + \lambda_\sigma\IB(R^2_\sigma) + 8 \lambda_\sigma\IB(R^2_a)\Bigr] \fineq{,} \\
		\begin{split}
			M^2_\pi(\sigCl) & = m^2_\pi(\sigCl) + \frac{T^2}{4\pi^2} \Bigl[ (\lambda_\sigma+2\lambda_a) \IB(R^2_\etap) + 5(2\lambda_\sigma+\lambda_a)\IB(R^2_\pi) \\
			& \hspace{7.5em} + \lambda_\sigma\IB(R^2_\sigma) + (8\lambda_\sigma+9\lambda_a) \IB(R^2_a) \Bigr] \fineq{,}
		\end{split}
	\end{align}
	\label{eq:CJT:gapLSM}%
\end{subequations}
where we have defined \mbox{$R_i\equiv R_i(\sigCl):=M_i(\sigCl)/T$}, while the field-dependent tree-level meson masses $m_i(\sigCl)$ can be found in \cref{eq:model:lsm:treeLevelMasses}.
Note that, following Refs.~\cite{Lenaghan2000a,Roder2003,Tsumura2017}, we consistently ignore vacuum contributions in evaluating loop integrals throughout all LSM-related calculations.
The remaining thermal parts, $\JB$ and $\IB$, of the relevant functions can be found in \cref{eq:integrals:thermal_int}.

For a given field value and at a given temperature, the system in \cref{eq:CJT:gapLSM} can be solved for the four effective masses $M_i$, which, in turn, define the full, field-dependent meson propagators according to the ansatz in \cref{eq:cjt:ansatz}.
The sought-after finite-temperature effective potential can then be computed via \cref{eq:cjt:Veff}, which eventually leads to the final expression quoted in \cref{eq:model:lsm:Veff}.

\section{Model parameters}
\label{app:parameters}
\noindent
In \cref{tab:app:parameters}, we collect our specific choices for the free parameters of the (P)NJL and linear sigma models, which correspond to the benchmark mass spectra listed in \cref{tab:pt:benchmark} and used throughout the paper.
How to calculate the meson masses as well as the pion decay constant for a given set of parameters is explained in the main text, namely around \cref{eq:model:njl:fPi} for the (P)NJL model and at the very end of \cref{sec:model:lsm} for the linear sigma model.

Note that the scaled-up benchmark points discussed in \cref{sec:comp} are obtained by rescaling the model parameters of \cref{tab:app:parameters} with $\xi$ according to their mass dimension, \eg \mbox{$G\to\xi^{-2}G$} etc., where the dimensionless factor $\xi$ was introduced in \cref{eq:results:rescaleMasses}.

\begin{table}[h]
	\centering
	\sisetup{round-mode=places}
	\renewcommand{\arraystretch}{1.4}
	\begin{tabular}{c|S[table-format=1.2,round-precision=2]S[table-format=-3.2,round-precision=2]|S[table-format=-6.0,round-precision=0]S[table-format=2.1,round-precision=1]S[table-format=2.1,round-precision=1]S[table-format=4.0,round-precision=0]}
		\toprule
		\multirow{2}{*}{\symhspace{0.5em}{\shortstack{benchmark\\point}}} &
		\multicolumn{2}{c|}{{(P)NJL model}} &
		\multicolumn{4}{c}{{Linear Sigma Model}} \\
		& {\symhspace{0.3em}{$G$ [GeV$^{-2}$]}}
		& {\symhspace{0.3em}{$G_D$ [GeV$^{-5}$]}}
		& {\symhspace{0.3em}{$m^2$ [\si{MeV^2}]}}
		& {\symhspace{0.9em}{$\lambda_\sigma$}}
		& {\symhspace{0.9em}{$\lambda_a$}}
		& {\symhspace{0.3em}{$c$ [MeV]}} \\
		\colrule
		A & 3.84 & -90.65 & -4209 & 16.76 & 12.91 & 2369 \\
		B & 3.988 & -106.445 & 4848 & 25.54 & 15.22 & 4091 \\
		C & 4.0 & -60.0 & 24371.2 & 14.84 & 13.99 & 1196.075 \\
		D & 5.0 & -60.0 & 192722.186 & 33.25 & 25.14 & 2176.0  \\
		\botrule
	\end{tabular}
	\caption{Model parameters corresponding to the mass spectra listed in \cref{tab:pt:benchmark}. The UV cutoff scale of the (P)NJL model is fixed to \mbox{$\Lambda = \SI{930}{MeV}$} for all benchmark points.}
	\label{tab:app:parameters}
\end{table}

\bibliographystyle{JHEP}
\bibliography{refsQCD}


\end{document}